\documentclass[PAPER, atlasdraft=false, UKenglish, cernpreprint, texmf, orcidlogo]{atlasdoc}
\usepackage{atlaspackage}
\usepackage{atlasbiblatex}
\usepackage{atlascontribute}

\usepackage{atlasphysics}
\graphicspath{{logos/}{figures/}}

\AtlasTitle{Calibration of the jet energy scale and resolution of small-radius jets using semileptonic $t\bar{t}$ events with the ATLAS detector}

\AtlasAbstract{%
A measurement of correction factors for the hadronic jet energy scale and resolution in the ATLAS detector is presented. These correction factors account for differences between simulated and observed data. They are obtained by analysing a selection of top quark events collected in proton--proton collisions by ATLAS between the years 2015 and 2018 at a centre-of-mass energy $\sqrt{s} = 13$~\tev as well as in 2022 and 2023 at $\sqrt{s} = 13.6$~\tev.
The impact of different jet energy scale or resolution corrections on the reconstructed mass of the hadronically decaying $W$ boson from top-quark decays in simulation is quantified.
The correction factors are extracted from a fit to the parameterised reconstructed $W$-boson mass distribution to data.
The energy scale and resolution corrections are measured as a function of the jet transverse momentum between 20 and 200~\gev and absolute pseudorapidity less than 0.8.
The uncertainties in the energy scale range from about 0.93\% to about 1.7\% for jets between 35 and 200~\gev, while for the energy resolution the uncertainties range from about 14\% to 28\%.
The method presented will be used in conjunction with other techniques to further improve ATLAS jet energy scale and resolution precision.
}

\author{The ATLAS Collaboration}

\AtlasRefCode{JETM-2024-05}
\PreprintIdNumber{CERN-EP-2025-262}

\AtlasJournal{EPJC}
\AtlasJournalRef{Eur. Phys. J. C 86 (2026) 943}
\AtlasDOI{10.1140/epjc/s10052-026-15692-2}

\AtlasCoverEgroupAnalysisTeam{atlas-jetm-2024-05-analysis-team@cern.ch}

\AtlasNote{JETM-2024-05}

\hypersetup{pdftitle={ATLAS document},pdfauthor={The ATLAS Collaboration}}

\begin{document}

\maketitle

\section{Introduction}

\newcommand{\AtlasCoordFootnote}{%
ATLAS uses a right-handed coordinate system with its origin at the nominal interaction point (IP)
in the centre of the detector and the \(z\)-axis along the beam pipe.
The \(x\)-axis points from the IP to the centre of the LHC ring,
and the \(y\)-axis points upwards.
Polar coordinates \((r,\phi)\) are used in the transverse plane,
\(\phi\) being the azimuthal angle around the \(z\)-axis.
The pseudorapidity is defined in terms of the polar angle \(\theta\) as \(\eta = -\ln \tan(\theta/2)\) and is equal to the rapidity
$ y = \frac{1}{2} \ln \left( \frac{E + p_z}{E - p_z} \right) $ in the relativistic limit.
Angular distance is measured in units of \(\Delta R \equiv \sqrt{(\Delta y)^{2} + (\Delta\phi)^{2}}\).}

In quantum chromodynamics (QCD), the theory of strong interactions, the quarks and gluons carry colour charge. Because of the confinement properties of QCD, they cannot be detected directly but instead form collimated sprays of hadrons, called jets.
Jets are produced abundantly at the Large Hadron Collider (LHC)~\cite{Evans:2008zzb} and are an important signature for many processes.
The measurement of the jet energy plays a vital role in all jet-based analyses with the ATLAS detector, with the uncertainty in the energy calibrations often being one of the largest uncertainties in precision measurements~\cite{STDM-2020-04,STDM-2019-02,TOPQ-2018-07,TOPQ-2022-24,TOPQ-2018-23}.

In ATLAS, the jet energy is evaluated through a series of corrections starting from the correction to the energy of the individual jet's constituents~\cite{PERF-2014-02,PERF-2015-09,PERF-2016-04,JETM-2018-05,JETM-2022-01}.
The calibration of the average jet transverse momentum~(\pt) is called the `jet energy scale' (JES) and the calibration of the \pt resolution is called the `jet energy resolution' (JER).
The jet energy corrections are based on Monte Carlo (MC) simulations modelling the detector response and on assessing \textit{in situ} how well the MC simulations describe data.
The corrections ensure that the reconstructed jet~\pt equals, on average, the~\pt of the corresponding particle-level jet, i.e. the event generator output after the hadronisation and before the detector simulation.
Several effects are corrected including the non-compensating nature of the ATLAS calorimeter (i.e. a lower response to hadrons than to electrons or photons), energy deposits outside jets and energy losses in dead material before and in-between the calorimeters.
The jet response is usually expressed as a function of variables, such as the energy fraction in the hadronic calorimeter or the number of charged-particle tracks associated to a jet, that are sensitive to jet fragmentation.

A subsequent set of \textit{in situ} jet calibrations are applied to ensure that the jet~\pt evaluated in data matches the simulation.
These corrections are obtained by exploiting the \pt-balance in certain event topologies where the jet recoils against a well-measured object like a photon or a $Z$ boson decaying into electron or muon pairs ($Z/\gamma$+jet), dijet or multijet events, and by comparing the \pt-balance in simulation with that in data.

The `$Z/\gamma$+jet' method was also used for JER measurements in previous ATLAS calibrations~\cite{PERF-2014-02}. Currently, ATLAS JER measurements rely on two methods.
Firstly, the `dijet-balance' method~\cite{JETM-2018-05}, where the scalar balance of the \pt of the jets in dijet events, is exploited.
Secondly, the `random cones' technique~\cite{JETM-2018-05}, in which energy deposits in the calorimeter are summed in circular areas analogous to the jet area for \antikt jets~\cite{Cacciari:2008gp}, providing the best precision for low~\pt jets.

This paper focuses on a complementary technique to improve the current \textit{in situ} calibration for the JES and JER by measuring jets in top-quark pair-production~(\ttbar) events, where one of the top-quark decays includes a lepton from the $W$ boson (semileptonic \ttbar events).
Semileptonic \ttbar events are chosen due to the presence of a hadronically decaying $W$ boson whose lineshape distribution is sensitive to JES and JER\@.
Additionally, semileptonic \ttbar events contain a high-momentum charged lepton that provides a clear signal for the trigger to accept the events.
Finally, the \ttbar production cross-section is large at the LHC and the decay products provide a signature with an isolated lepton and $b$-jet in the final state that leads to an experimental signature with minimal background.
The ATLAS Collaboration previously used \ttbar events as a cross-check of the JES calibration~\cite{PERF-2012-01}.
Moreover, top-quark mass measurements performed at Tevatron and LHC experiments exploited the $W$-boson lineshape distribution to control the JES uncertainties~\cite{PhysRevD.95.112004,TOPQ-2019-13,atlas2014combination,TOPQ-2013-02,TOPQ-2017-03,TOPQ-2022-24}, demonstrating the potential of $W$-boson mass distribution as a powerful probe of the jet response.
However, the obtained corrections to the JES in the previous top-quark mass measurements are not considered universal as they are not binned in jet~\pt and jet $\eta$ but are specific to each measurement's selections, and are used only within their respective measurements.
This paper presents the measurement of the JES corrections as a function of jet~$\pt$ and jet $|\eta|$ making the calibration universal and not relying on the kinematics of the calibration dataset.
The selection of the events in the presented measurement is orthogonal to the selection with the established jet calibration techniques.
The dominant uncertainties of the presented measurement originate from the uncertainty in the modelling of the \ttbar events, making the systematic uncertainties decorrelated from the standard calibration techniques.
Additionally, the jets from the $W$-boson decay are quark-initiated, compared to mostly gluon-initiated jets from the established JER calibration techniques providing complementary information about jet calibrations.

This paper presents a measurement of the correction factors between simulation and data for the JES and the JER, targeting jets clustered with the \antikt algorithm with a radius parameter $R = 0.4$ using the proton--proton collision data collected by the ATLAS detector in Run~2, i.e. between the years 2015 and 2018 at a centre-of-mass energy of $\sqs = 13~\tev$ and, separately, for data collected in Run~3 in the years 2022 and 2023 at $\sqs = 13.6~\tev$.

The forward-folding formula~\cite{ATLAS-CONF-2016-008} is applied to individual jets to generate templates representing different JES or JER assumptions.
The forward-folding technique is used because of its better precision compared to the background suppression technique~\cite{ATLAS-CONF-2016-008}.
The two jets originating from the hadronically decaying $W$ boson are identified and the templates for the invariant mass of the two jets are constructed.
Only central jets in the pseudorapidity region of $|\eta| < 0.8$\footnote{\AtlasCoordFootnote} are considered to facilitate the future combination with other \textit{in situ} measurements.
A likelihood fit of the generated templates to the observed data is performed to find the optimal values for the JES and JER parameters together with their uncertainties.

\FloatBarrier


%
\section{ATLAS detector}

The ATLAS detector~\cite{PERF-2007-01,GENR-2019-02} at the LHC covers nearly the entire solid angle around the collision point.
It consists of an inner tracking detector surrounded by a thin superconducting solenoid, electromagnetic and hadronic calorimeters,
and a muon spectrometer incorporating three large superconducting air-core toroidal magnets.

The inner-detector system (ID) is immersed in a \qty{2}{\tesla} axial magnetic field
and provides charged-particle tracking in the range of \(|\eta| < 2.5\).
The high-granularity silicon pixel detector covers the vertex region and typically provides four measurements per track,
the first hit generally being in the insertable B-layer (IBL).
It is followed by the semiconductor tracker (SCT), which usually provides eight measurements per track.
These silicon detectors are complemented by the transition radiation tracker (TRT),
which enables radially extended track reconstruction up to \(|\eta| = 2.0\).
The TRT also provides electron identification information
based on the fraction of hits (typically 30 in total) above a higher energy-deposit threshold corresponding to transition radiation.

The calorimeter system covers the pseudorapidity range \(|\eta| < 4.9\).
Within the region \(|\eta|< 3.2\), electromagnetic calorimetry is provided by barrel and
endcap high-granularity lead/liquid-argon (LAr) calorimeters,
with an additional thin LAr presampler covering \(|\eta| < 1.8\)
to correct for energy loss in material upstream of the calorimeters.
Hadronic calorimetry is provided by the steel/scintillator-tile calorimeter,
segmented into three barrel structures within \(|\eta| < 1.7\), and two copper/LAr hadronic endcap calorimeters.
The solid angle coverage is completed with forward copper/LAr and tungsten/LAr calorimeter modules
optimised for electromagnetic and hadronic energy measurements, respectively.

The muon spectrometer (MS) comprises separate trigger and
high-precision tracking chambers measuring the deflection of muons in a magnetic field generated by the superconducting air-core toroidal magnets.
The field integral of the toroids ranges between \num{2.0} and \qty{6.0}{\tesla\metre}
across most of the detector.
Three layers of precision chambers, each consisting of layers of monitored drift tubes, cover the region \(|\eta| < 2.7\).
These were complemented in the innermost layer of the endcap region by cathode-strip chambers in Run~2,
which were replaced in Run~3 by layers of small-strip thin-gap chambers and Micromegas chambers, both providing precision tracking in the region \(1.3 < |\eta| < 2.7\).
The muon trigger system covers the range \(|\eta| < 2.4\) with resistive-plate chambers in the barrel region, thin-gap chambers in the endcap regions,
and, in Run~3, the small-strip thin-gap chambers and Micromegas chambers in the innermost layer of the endcap.

The luminosity is measured mainly by the LUCID--2 detector~\cite{LUCID2} that records Cherenkov light produced in the quartz windows of photomultipliers located close to the beampipe.

Events were selected by the first-level trigger system implemented in custom hardware,
followed by selections made by algorithms implemented in software in the high-level trigger~\cite{TRIG-2016-01,TRIG-2022-01}.
The first-level trigger accepted events from the \qty{40}{\MHz} bunch crossings at a rate close to \qty{100}{\kHz},
which the high-level trigger further reduced in order to record complete events to disk at about \qty{1.25}{\kHz} and \qty{3}{\kHz} in Run~2 and Run~3, respectively.

The Run~3 detector configuration benefits from several upgrades compared with that of Run~2 to maintain high detector performance at the higher pile-up levels of Run~3.
The improvements include a new innermost layer of the muon spectrometer in the endcap region, which provides higher redundancy and a large reduction in fake muon triggers.
The trigger system also benefits from new digital electronics readout of the LAr calorimeters with significantly increased granularity.
Other updates and further details are provided in Ref.~\cite{GENR-2019-02}.

A software suite~\cite{SOFT-2022-02} is used in data simulation, in the reconstruction
and analysis of real and simulated data, in detector operations, and in the trigger and data acquisition
systems of the experiment.

\FloatBarrier


%
\section{Data and Monte Carlo samples}

This analysis uses two data samples collected by ATLAS\@.
Only events recorded when LHC beams were stable and all ATLAS detector systems were operational are selected~\cite{DAPR-2018-01}.
The first sample consists of the full LHC Run~2 data collected in 13~\tev proton--proton ($pp$) collisions during 2015–2018, corresponding to an integrated luminosity of $140~\ifb$~\cite{DAPR-2021-01}.
The second consists of LHC Run~3 $pp$ data collected during 2022 and 2023, corresponding to an integrated luminosity of $52~\ifb$.
The uncertainty in the total integrated luminosity is 0.83\% for the Run~2 data-taking and 2\% for the Run~3 data-taking~\cite{DAPR-2021-01, ATL-DAPR-PUB-2024-001}.
The mean average number of interactions per bunch crossing was 33.7 for Run~2 and 42.5 and 51.2 during Run 3 data-taking in 2022 and 2023 respectively.

Only data events selected by the single-electron or single-muon trigger~\cite{TRIG-2018-01,TRIG-2018-05} are used in the presented analysis.
The electron trigger reached the efficiency plateau region for electrons with reconstructed $\pt > 25$~\gev in 2015 and for $\pt > 27$~\gev in 2016–2023.
The corresponding thresholds for the muon trigger were 21~\gev in 2015, 27~\gev in 2016--2018 and 25~\gev in 2022--2023.

The top-quark mass was set to $m_{t} = 172.5~\gev$ in all simulated processes, unless stated otherwise.
The decays of bottom and charm hadrons were performed by \EVTGEN[1.6.0]~\cite{Lange:2001uf} in the Run~2 simulation and by \EVTGEN[2.1.1] in the Run~3 simulation.

The production of \ttbar events was simulated using the \POWHEGBOX[v2]~\cite{Frixione:2007nw,Nason:2004rx,Frixione:2007vw,Alioli:2010xd} generator at next-to-leading order (NLO) with the \NNPDF[3.0nlo]~\cite{Ball:2014uwa} parton distribution function (PDF) set and the \hdamp parameter\footnote{The \hdamp parameter is a resummation damping factor and one of the parameters that controls the matching of \POWHEG matrix elements to the parton shower. It essentially regulates the high-\pT radiation against which the \ttbar system recoils.} set to $1.5\,m_{t}$~\cite{ATL-PHYS-PUB-2016-020}.
The \ttbar sample was normalised to the cross-section prediction at next-to-next-to-leading order (NNLO)
in QCD including the resummation of next-to-next-to-leading-logarithmic (NNLL) soft-gluon terms calculated using
\TOPpp[2.0]~\cite{Beneke:2011mq,Cacciari:2011hy,Baernreuther:2012ws,Czakon:2012zr,Czakon:2012pz,Czakon:2013goa,Czakon:2011xx}.

The associated production of top quarks with \(W\) bosons (\(tW\)) was simulated by the \POWHEGBOX[v2]~\cite{Re:2010bp,Nason:2004rx,Frixione:2007vw,Alioli:2010xd} generator at NLO in QCD using the five-flavour scheme and the \NNPDF[3.0nlo] set of PDFs.
The functional form of the renormalisation and factorisation scales was set to a dynamic scale with $\mu = H_\text{T}/2 = (m_{\text{T,top}} + m_{\text{T},W} + p_{\text{T,j1}})/2$, where $m_{\text{T,top}}$ and $m_{\text{T},W}$ are the transverse masses of top quark and $W$ boson, respectively, and $p_{\text{T,j1}}$ is the transverse momentum of the leading jet.
The diagram removal scheme~\cite{Frixione:2008yi} was used to remove interference and overlap with \ttbar production.
The inclusive cross-section was corrected to the theory prediction calculated at NLO in QCD with NNLL soft-gluon corrections~\cite{Kidonakis:2010ux,Kidonakis:2013zqa}.
The $tW$ process dominates the contribution of the single-top processes in the analysis selection described below.

Single-top \(t\)-channel and \(s\)-channel production was simulated using the \POWHEGBOX[v2]~\cite{Frederix:2012dh,Alioli:2009je} generator at NLO in QCD using the four-flavour (five-flavour) scheme for the \(t\)-channel (\(s\)-channel) and the corresponding \NNPDF[3.0nlo] set of PDFs.

For all nominal simulated \ttbar and single-top samples, the events were interfaced to \PYTHIA[8.230]~\cite{Sjostrand:2014zea} for the Run~2 simulation and interfaced to \PYTHIA[8.308] for the Run~3 simulation to model the parton shower, hadronisation, and underlying event, with parameters set according to the A14 set of tuned parameters (tune)~\cite{ATL-PHYS-PUB-2014-021} and using the \NNPDF[2.3lo] set of PDFs~\cite{Ball:2012cx}.
Several alternative MC generators are used to estimate the uncertainty in the modelling of the \ttbar and $tW$ events as described in Section~\ref{sec:systematics}.

The production of $(Z/W)$+jets events was simulated with the \SHERPA[2.2.11] and \SHERPA[2.2.14]~\cite{Bothmann:2019yzt} generators using NLO matrix elements (ME) for up to two partons, and leading-order (LO) matrix elements
for up to four (five) partons in the Run~2 (Run~3) simulation calculated with the Comix~\cite{Gleisberg:2008fv} and \OPENLOOPS~\cite{Buccioni:2019sur,Cascioli:2011va,Denner:2016kdg} libraries.
They were matched to the \SHERPA parton shower~\cite{Schumann:2007mg} using the \MEPSatNLO prescription~\cite{Hoeche:2011fd,Hoeche:2012yf,Catani:2001cc,Hoeche:2009rj} with the set of tuned parameters developed by the \SHERPA authors.
The \NNPDF[3.0nnlo] set of PDFs~\cite{Ball:2014uwa} was used and the samples were normalised to an NNLO prediction~\cite{Anastasiou:2003ds}.

Samples of diboson final states (\(VV\)) were simulated with \SHERPA[2.2.11] or \SHERPA[2.2.12] for the Run~2 simulation, and \SHERPA[2.2.14] or \SHERPA[2.2.16] for the Run~3 simulation, depending on the process, including off-shell effects and Higgs boson contributions, where appropriate.
Fully leptonic final states and semileptonic final states, where one boson decays leptonically and the other hadronically, were simulated using matrix elements at NLO accuracy in QCD for up to one additional parton and at LO accuracy for up to three additional parton emissions.

Simulated \ttbar events with both $W$ bosons decaying hadronically are used to estimate the contribution of the fake and non-prompt lepton background.
The flavour-tagging requirements with the kinematic reconstruction presented in Section~\ref{sec:event_selection} significantly suppress the other sources of the fake and non-prompt backgrounds and the contribution is expected to be dominated by the \ttbar events.

The effect of pile-up was modelled by overlaying~\cite{SIMU-2020-01} the simulated hard-scattering event with inelastic \(pp\) events generated from a mix of \EPOS[2.0.1.4]~\cite{Werner:2005jf} and \PYTHIA[8.308]~\cite{Bierlich:2022pfr}.
The \EPOS events were generated with the \EPOS LHC tune~\cite{Pierog:2013ria} and the \PYTHIA events with the A3 tune~\cite{ATL-PHYS-PUB-2016-017} and the \NNPDF[2.3lo] set of PDFs.
\PYTHIA pile-up events include either a high \pt jet, a prompt photon, or a lepton from a $B$-hadron decay, while \EPOS was filtered to simulate all remaining pile-up events in the overlay sample.
The individual simulations were first reweighted to ensure a smooth connection across jet \pt, then the combination reweighted to match the distribution of the actual number of interactions per bunch crossing measured in data.

The ATLAS detector response was simulated using the \GEANT toolkit~\cite{Agostinelli:2002hh} with the full simulation of the ATLAS detector~\cite{SOFT-2010-01} for most of the samples.
For some of the systematic variations affecting the measurement of the JES and JER correction factors, a fast detector simulation based on a parametric description of the calorimeter showers~\cite{SIMU-2018-04} was used.
The samples with fast detector simulations are always compared only to nominal samples with the same fast detector simulations.

\FloatBarrier


%
\section{Object definitions and forward-folding method}
\label{sec:objects}

Electrons are reconstructed from energy deposits in the electromagnetic calorimeter matched to a track in the ID\@.
The electrons are required to satisfy the `tight' likelihood-based identification working point and to be well isolated~\cite{PERF-2017-01,EGAM-2018-01}.
These electrons are required to have $\pt > 27~\gev$ and $|\eta| < 2.47$ while excluding the transition region between the barrel and endcap calorimeters ($1.37 < |\eta| < 1.52$).
Muons are reconstructed from a track in the ID and a matched track from the MS.
The muons are required to satisfy the `medium' identification working point and to be well isolated~\cite{MUON-2022-01,MUON-2018-03}.
The muons are required to have $\pt > 27~\gev$ and $|\eta| < 2.5$.
The identification, isolation and trigger efficiencies of the electrons and muons are corrected in simulation via scale factors to match the efficiencies measured in data~\cite{EGAM-2018-01,TRIG-2018-05,TRIG-2018-01}.

Electron (muon) candidates are matched to the reconstructed position of the main interaction point, the primary vertex~\cite{ATL-PHYS-PUB-2015-026}, by requiring that the significance of their transverse impact parameter, $d_0$,\footnote{The transverse impact parameter, $d_0$, is defined in the $x$--$y$ plane as the distance of closest approach of the track to the beamline.} satisfies $|d_0/\sigma(d_0)| < 5~(3)$, where $\sigma(d_0)$ is the measured uncertainty in $d_0$, and by requiring that their longitudinal impact parameter, $z_0$,\footnote{The longitudinal impact parameter, $z_0$, is defined as the distance in $z$ between the primary vertex and the point on the track used to evaluate $d_0$.} satisfies $|z_0 \sin{\theta}| < 0.5$~mm.

The jets are reconstructed from particle flow objects~\cite{PERF-2015-09} using the \antikt clustering algorithm with a radius parameter $R = 0.4$, implemented in the \textsc{FastJet} software package~\cite{Fastjet}.
Only jets that have $\pt > 15~\gev$ and $|\eta| < 2.5$ are considered for the selection, while only jets with $\pt > 20~\gev$ and $|\eta| < 0.8$ are used for the JES and JER correction measurement.
To suppress jets originating from pile-up, jets with $\pt < 60~\gev$ are required to satisfy a neural-network-based jet vertex tagger (NNJVT) discriminant, a successor to the jet vertex tagger algorithm~\cite{PERF-2014-03}.
The reconstructed jets are calibrated in two different ways for use in this measurement, once with no \textit{in situ} corrections applied for examining the JES and once with the currently used \textit{in situ} corrections~\cite{JETM-2018-05} for examining the JER\@.  In both cases, the MC-based JES correction is applied.
Using these two separate calibrations for the different studies ensures that the results can be compared with other \textit{in situ} calibration techniques, and that the best-available calibration is used as the basis for each step.

Jets that include $B$-hadrons ($b$-tagged jets) are identified by the GN2 flavour tagging algorithm~\cite{FTAG-2023-05} with a working point corresponding to an efficiency of $85\%$ for selecting a true $b$-jet in a \ttbar reference sample.
Scale factors are applied to simulation to match the efficiency of the $b$-tagging measured in data~\cite{FTAG-2018-01, FTAG-2019-02, FTAG-2020-08}.

The missing transverse momentum is defined as the negative vector sum of the transverse momenta of all selected and calibrated physics objects, including a term to account for the momenta of soft particles that are not associated with any of the selected objects~\cite{JETM-2020-03}.
This soft term is calculated from ID tracks matched to the primary vertex.
The magnitude of the missing transverse momentum vector is denoted by \met.

To avoid any double-counting of physics objects, a dedicated sequential overlap-removal procedure is applied to resolve the ambiguities.
Jets within $\Delta R = 0.2$ of an electron or muon are discarded and electrons with $0.2 < \Delta R < 0.4$ to the remaining jets are rejected.
Jets with fewer than three tracks are removed if they overlap with a muon within $\Delta R = 0.4$.

For the forward-folding procedure~\cite{ATLAS-CONF-2016-008, ATLAS-CONF-2016-035, ATLAS-CONF-2017-063}, particle-level objects need to be identified, using the MC particle-level record in simulation.
Particle-level leptons are selected as leptons originating from the decay of a $W$ or $Z$ boson.
The four-momentum of an electron or muon is summed with the four-momenta of all radiated photons within a cone of size $\Delta R = 0.1$ around its direction, excluding photons from hadron decays.
The particle-level jets are reconstructed with the \antikt algorithm with a radius parameter of $R=0.4$, using all stable particles ($c\tau > 10~\text{mm}$) except for the selected electrons, muons, and photons used in the definition of the selected leptons, and neutrinos originating from the $Z$ boson or $W$ bosons.
The particle-level jets are required to have $\pt > 10~\gev$ and $|\eta| < 2.5$.

The forward-folding method is used to simulate the changes to the~\pt of the selected jets due to alternative assumptions on the values of the JES or JER and it is implemented in the \textsc{FastFrames} software library~\cite{FastFrames}.
The method relies on matching the detector-level jets to the corresponding particle-level jets and changing the momenta of reconstructed jets using the formula

\begin{equation}
\pt^{\text{folded}}=s\cdot \pt^{\text{reco}}+(\pt^{\text{reco}} - \mathcal{R} \cdot \pt^{\text{gen}})\cdot(r-s),
\label{eq:FF}
\end{equation}

where $\pt^{\text{gen}}$ is the momentum of the generated particle-level jet matched to the reconstructed jet with transverse momentum $\pt^{\text{reco}}$ and $s$ ($r$) is the parameter representing the flavour-inclusive correction factor to the simulated JES (JER).
The correction factors from Eq.~(\ref{eq:FF}) are consistently applied to the jet's~\pt and its energy.
The jet energy response and resolution is averaged over $\eta$ in the relevant analysis region with jet $|\eta| < 0.8$ as the detector is approximately homogeneous in this narrow central region.
The forward-folding formula is used to generate templates, i.e. distributions representing various JES and JER correction factors with values ranging from $0.95$ to $1.05$ and $0.60$ to $1.40$, respectively, to cover the range of the expected corrections.
The parameter $\mathcal{R}$ represents the mode of the jet response distribution

\begin{equation*}
\mathcal{R} = \text{argmax} \left\{ \frac{\pt^{\text{reco}}}{\pt^{\text{gen}}} \right\},
\end{equation*}

estimated in simulation by fitting a Gaussian function around the peak position separately in several jet~\pt and $|\eta|$ bins, for different jet flavours and for LHC Run~2 and Run~3 simulations.
The following jet flavours are identified for the parameter $\mathcal{R}$ using the MC particle-level record: $b$-jet, $c$-jet, light-flavour ($u$, $d$, $s$)-jet and gluon-jet.
The $\mathcal{R}$ parameter is found to be close to unity in most of the kinematic regions and for most jet flavours.
The JES variations are generated while keeping the nominal JER assumption, i.e. setting the parameter $r$ in Eq.~(\ref{eq:FF}) to unity and vice-versa for the JER variations.
All further procedures are applied after a specific set of $r$ and $s$ parameters are applied.

For each reconstructed jet, the closest particle-level jet in angular distance is identified.
The reconstructed and particle-level jets are matched if the angular distance $\Delta R$ is smaller than 0.2 and there is only one particle-level jet satisfying the requirement for a given reconstructed jet.
Additionally, isolation is imposed on the matched particle-level jet by requiring that no other particle-level jet is found within $\Delta R = 0.5$.
Furthermore, to ensure the measured energy for a given reconstructed jet represents only the jet energy, no particle-level electrons or muons originating from on-shell $W$ or $Z$ bosons are allowed within $\Delta R = 0.5$ of the reconstructed jet.
In simulation, under these criteria, more than 90\% of the selected jets with $\pt < 150~\gev$ can be matched to the corresponding particle-level jets, while the fraction significantly drops for jets with $\pt > 150~\gev$ to about 25\% in the most extreme case, where most of the jets fail to match due to the isolation requirements of the particle-level jets.
Detector-level jets failing either the matching to the truth jet or truth-isolation requirements are not included in forward-folding, and they appear unmodified in the $W$-boson mass distributions.
All other jets, including jets with \pt above 200~\gev and $|\eta| > 0.8$ (but below $|\eta|= 2.5$) and jets that are not identified to come from the $W$-boson decay are forward-folded.
Changing the angular distance for the matching from $\Delta R < 0.2$ to $\Delta R < 0.3$ has no significant impact on the final results.

\FloatBarrier


%
\section{Event selection and $W$-boson mass reconstruction}
\label{sec:event_selection}

Events are required to have at least four jets with $\pt > 20~\gev$ and at least two jets need to be $b$-tagged and have $\pt > 30~\gev$.
Only events with exactly one electron or one muon with $\pt > 27~\gev$ are selected.
The lepton~\pt requirement is imposed to ensure the trigger efficiency is high.
Additionally, the events are required to have $\met > 30~\gev$.
These criteria are imposed on all events \emph{after} the application of the forward-folding procedure described in Section~\ref{sec:objects} for each folded set of jets, as the results of the folding can change the event selection decision.
To achieve this, the forward-folding technique is firstly applied to all jets in simulation for a looser selection of events with no \met requirements and with the jet~\pt selection lowered to 15~\gev.

The impact of the folding on \met is estimated by propagating the difference between the folded and original jet four-momenta using the following equation
\begin{equation*}
\vec{\pt}^{\text{\,miss, folded}} = \vec{\pt}^{\text{\,miss, original}} - \sum_{i} \left(\vec{\pt}^{\text{\,jet folded,i}} - \vec{\pt}^{\text{\,jet original,i}}\right),
\end{equation*}

with the missing transverse momentum vector $\vec{\pt}^{\text{miss}}$ and jet transverse momentum vector $\vec{\pt}^{\text{\,jet}}$ of jet $i$.

A simple permutational algorithm is used to identify the two jets from the hadronically decaying $W$ boson.
From all permutations of reconstructed jets, those that give the best agreement with the expected top-quark and $W$-boson masses---quantified by the $\chi^2$ value defined below---are selected

\begin{equation}
\chi^2 = \frac{(m_t^{\text{reco}} - m_t^{\text{gen}})^2}{\Sigma_t^2} + \frac{(m_W^{\text{reco}} - m_W^{\text{gen}})^2}{\Sigma_W^2}.
\label{eq:chi2}
\end{equation}

Here $m_t^{\text{reco}}$ represents the invariant mass of the three jet system from the top-quark decay, $m_{t}^{\text{gen}}$ represents a fixed constant for the top-quark mass set to 172.5~\gev and the constant $\Sigma_t$ represents the decay width of the top quark as well as the approximate experimental resolution and is set to 35~\gev.
A systematic uncertainty varying the top-quark mass is considered as described in Section~\ref{sec:systematics}.
Similarly, $m_W^{\text{reco}}$ represents the invariant mass of the two jet system from the $W$-boson decay, $m_{W}^{\text{gen}}$ represents the $W$-boson mass set to  80.38~\gev and $\Sigma_W$ represents the decay width and resolution of the $W$ boson and is set to 25~\gev.
The $\chi^2$ value is evaluated for permutations of up to six jets.
If there are more than six jets in an event, at least two $b$-tagged jets (ordered in~\pt) are always considered and then up to four other jets (ordered in~\pt) are considered.
For the permutations, within each top-quark combination, the jet not included in the $W$ boson must be $b$-tagged and permutations corresponding to the same top-quark or $W$-boson combinations are only considered once.
The permutation with the best (lowest) $\chi^2$ value is chosen.
Figure~\ref{fig:chi2} shows the distribution of the $\chi^2$ value for the best permutations in the Run~2 data sample without the JES \textit{in situ} corrections applied.
Good agreement between the prediction and data can be seen within the uncertainties.
To improve the matching purity, only events for which the best permutation has $\chi^2 < 1$ are considered.
This requirement removes about half of the pre-selected events satisfying all the previous criteria except the $\chi^2$ requirement.
Changing the parameters $\Sigma_t$ and $\Sigma_W$ in Eq.~(\ref{eq:chi2}) to 15 and 10~\GeV, respectively, and changing the $\chi^2$ selection to keep the same selection efficiency, does not impact the matching purity significantly.

Additionally, the reconstructed jets identified as coming from the hadronically decaying $W$ boson are required to be well isolated.
For the reconstructed jets, a different isolation procedure compared to the particle-level jets is employed.
A jet isolation variable $f_{\text{iso}}$ is defined by

\begin{equation*}
f_{\text{iso}} = \frac{\pt^{\text{iso}} - \rho A_{\text{iso}}}{\pt^{\text{jet}} - \rho A_{\text{jet}}},
\end{equation*}

where $\pt^{\text{iso}}$ represents the scalar sum of the~\pt of all inputs to the jet algorithm within an isolation annulus with inner and outer radii of 0.4 and 0.5. The ratio of momenta is calculated after subtracting the pile-up contribution, which is obtained by multiplying the jet active area $A_{\text{jet}}$ and the area of the isolation annulus $A_{\text{iso}} = \pi \cdot 0.5^2 - A_{\text{jet}}$ by the pile-up density $\rho$~\cite{Cacciari:2007fd}.
The jets from the $W$-boson decay are required to have $f_{\text{iso}} < 0.3$ to select well isolated jets.
An alternative isolation requirement based on the angular distance between the closest reconstructed jets is found to give consistent results with the isolation variable requirement.

Figure~\ref{fig:Wmass_insitu_off} illustrates the reconstructed $W$-boson mass distribution in Run~2 and Run~3 after applying all the selection criteria including the $\chi^2 < 1$ requirement in data without the \textit{in situ} jet calibration corrections applied.
There is good agreement between the prediction and data within the uncertainties.

\begin{figure}[ht]
\centering
\includegraphics[width=0.5\textwidth]{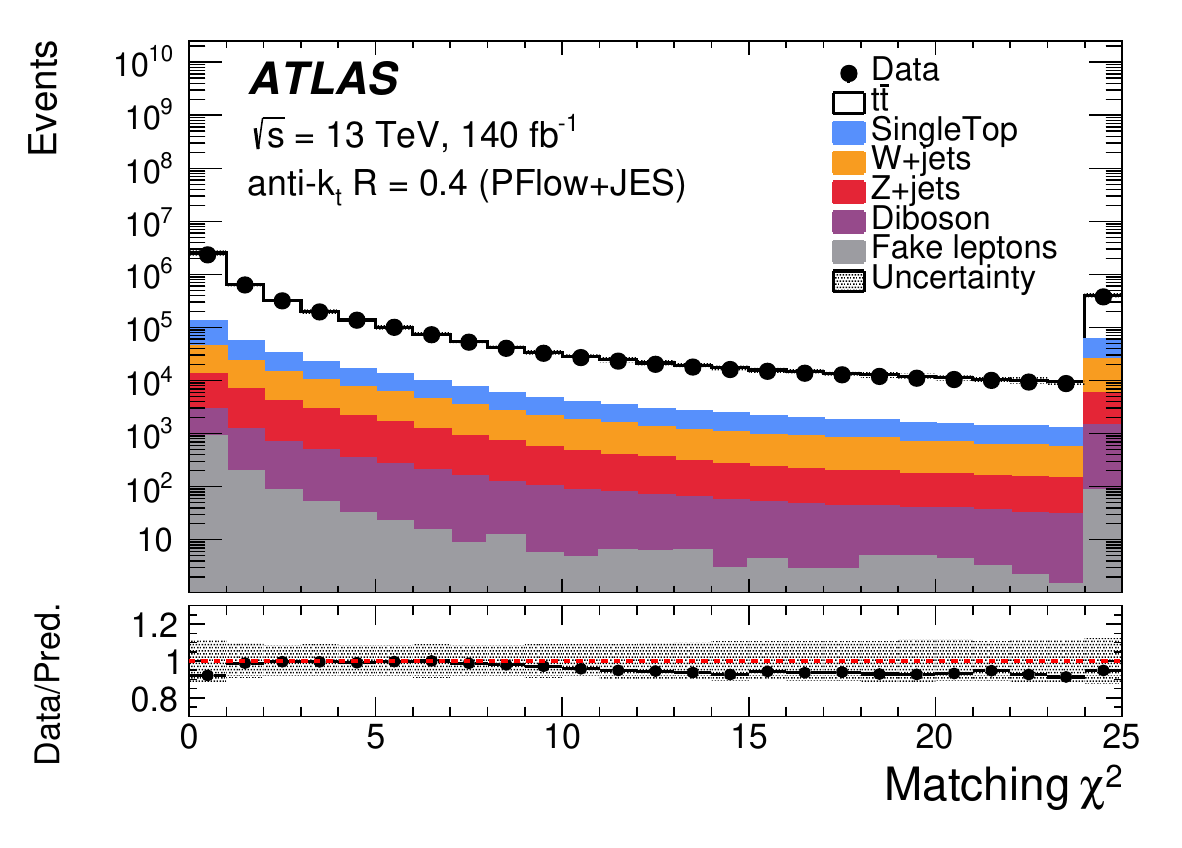}
\caption{The distribution of the best $\chi^2$ value for the considered permutations for matching reconstructed jets to jets from the top-quark decay in the Run~2 data. Data without the JES \textit{in situ} calibration (PFlow+JES) are shown as markers and compared to simulation from the signal and background processes. The uncertainty band includes all systematic uncertainties as described in Section~\ref{sec:systematics}. The bottom panel shows the ratio of the data to the prediction. The dashed line represents the ratio of one. The last bin contains the overflow events.}
\label{fig:chi2}
\end{figure}

\begin{figure}[ht]
\centering
\subfloat[]{\includegraphics[width=0.49\textwidth]{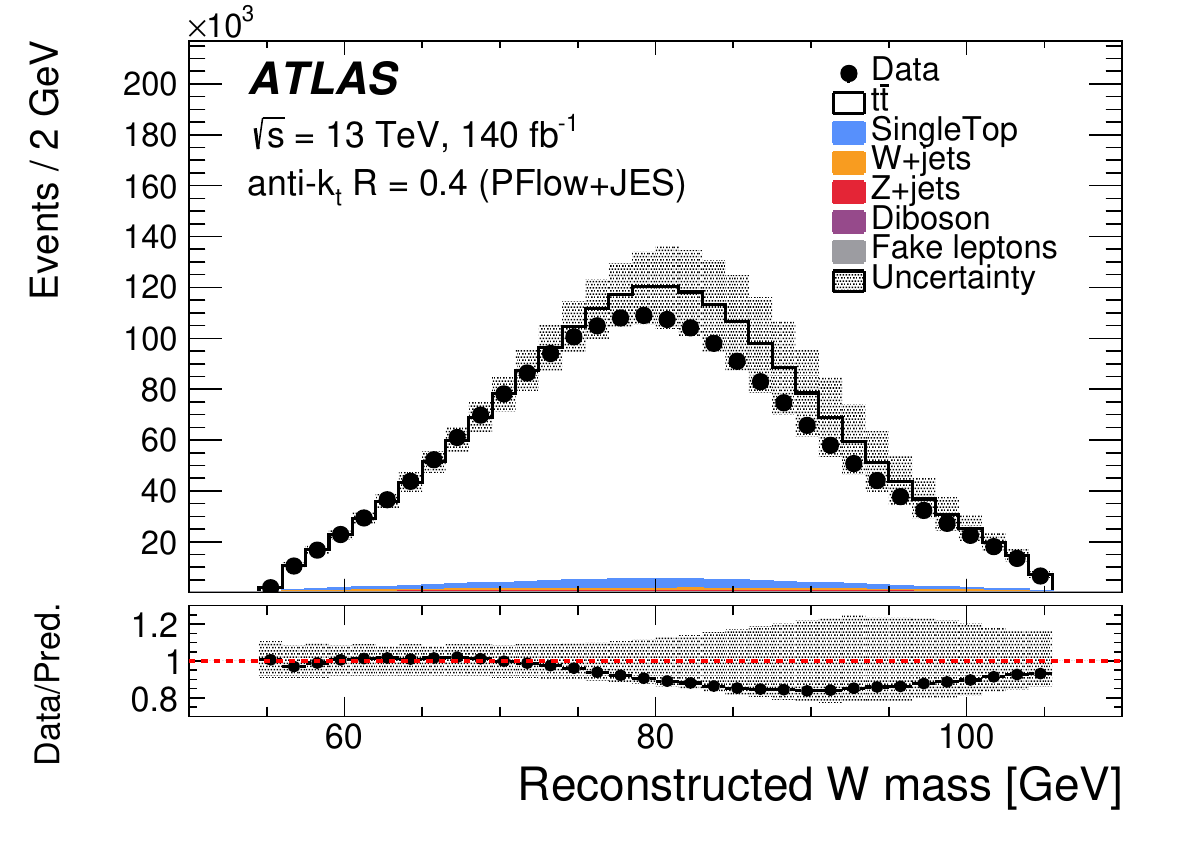}}
\subfloat[]{\includegraphics[width=0.49\textwidth]{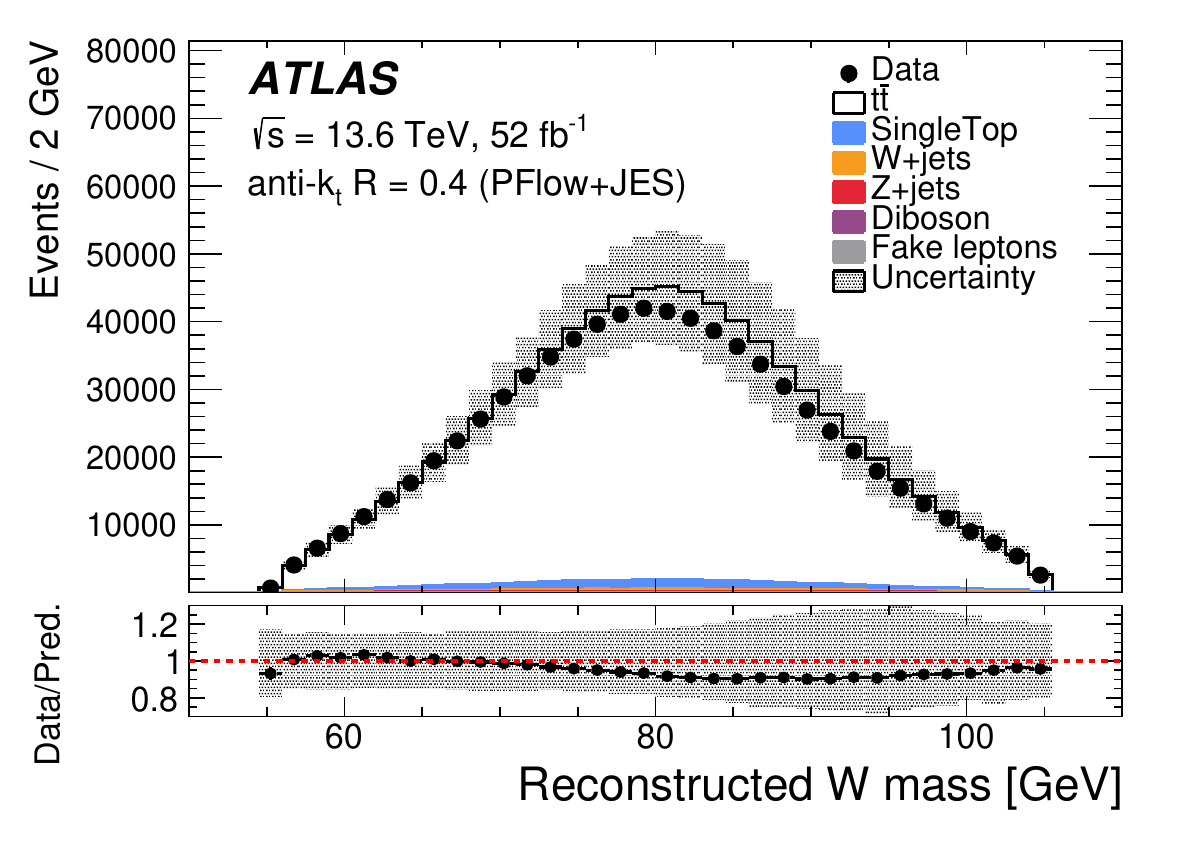}}
\caption{Reconstructed hadronically decaying $W$-boson mass after applying all selection criteria in the (a) Run~2 data and (b) Run~3 data without the JES \textit{in situ} corrections applied (PFlow+JES). The uncertainty band includes all systematic uncertainties. The bottom panels show the ratio of the data to the prediction. The dashed line represents the ratio of one.}
\label{fig:Wmass_insitu_off}
\end{figure}

Only events where both of the jets from the $W$-boson decay have a $\pt$ value in the range of 20--200~\gev, and $|\eta| < 0.8$ are analysed further.
As the JES and JER corrections are determined as a function of the jet~\pt, the distribution of the invariant mass of the two jets from the $W$-boson decay is analysed in multiple regions corresponding to bins in jet~\pt for the leading and subleading jets.
The \emph{diagonal} regions refer to events where jets fall into the same jet~\pt bin, while the remainder are referred to as \emph{off-diagonal} regions.
Figure~\ref{fig:Regions} illustrates the definition of the regions used for the measurement and the selected events in each of the regions in Run~2 data without \textit{in situ} corrections.
Due to the kinematic restrictions of the $W$-boson decay, the diagonal region for the jet~\pt range of 20--35~\gev is not considered further in the analysis.
Figure~\ref{fig:template_50_70} displays the reconstructed $W$-boson mass for various JES and JER assumptions in the diagonal region with jet~\pt between 50 and 70~\gev in the Run~2 data.
It can be seen that the JES assumptions mostly shift the distribution, while the JER assumptions mostly affect the width of the distribution.

\begin{figure}[ht]
\centering
\includegraphics[width=0.5\textwidth]{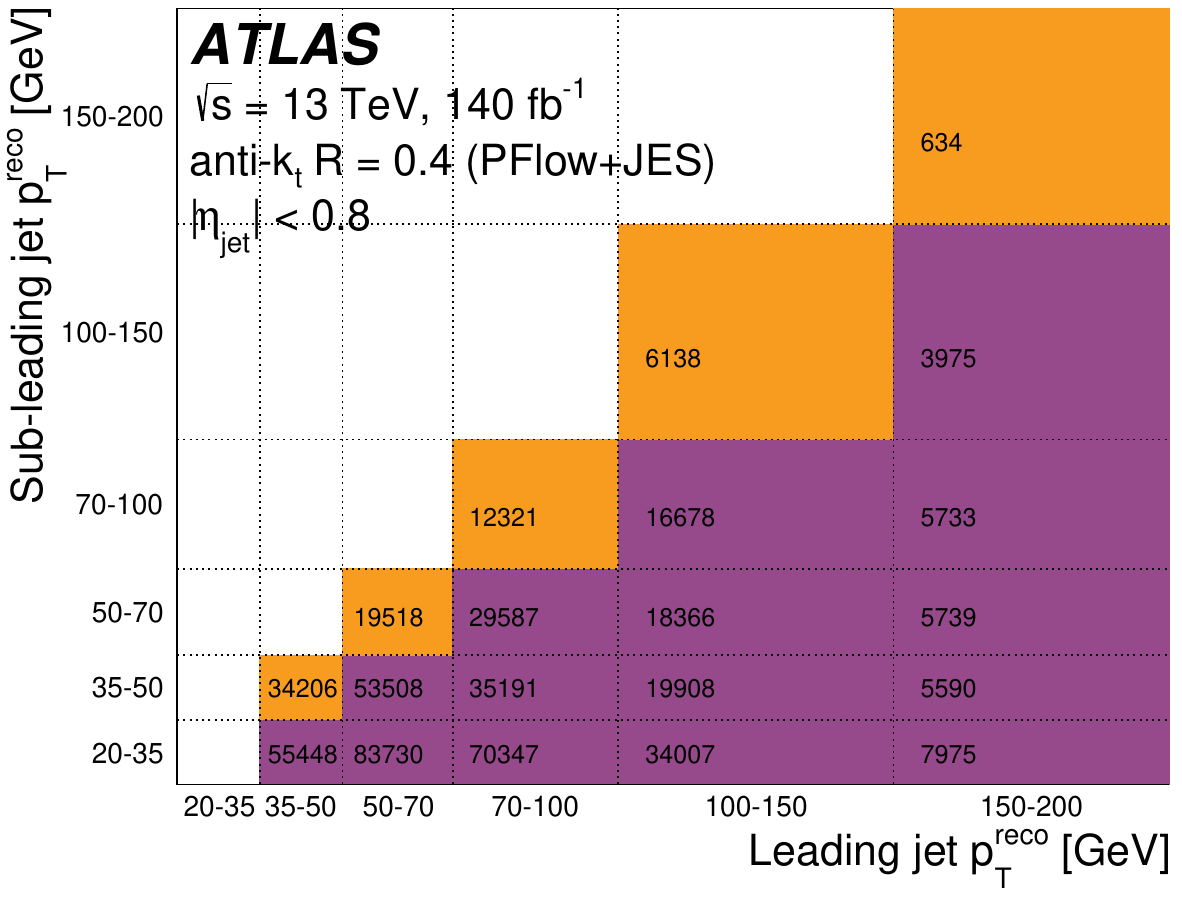}
\caption{Definition of the analysis regions based on reconstructed \pt and $\eta$ of the two jets identified to originate from the $W$-boson decay. The two jets from $W$ boson are required to have $|\eta|$ < 0.8. The regions shown on the diagonal are the regions where both jets from the $W$ boson fall into the same reconstructed \pt bin, off-diagonal are the regions where the two jets fall into different reconstructed \pt bins. The diagonal region with jets having a reconstructed~\pt between 20 and 35~\gev is not used due to kinematic restrictions. The numbers in the bins represent the number of selected events in Run~2 data without \textit{in situ} corrections.}
\label{fig:Regions}
\end{figure}

\begin{figure}[ht]
\centering
\subfloat[]{\includegraphics[width=0.80\textwidth]{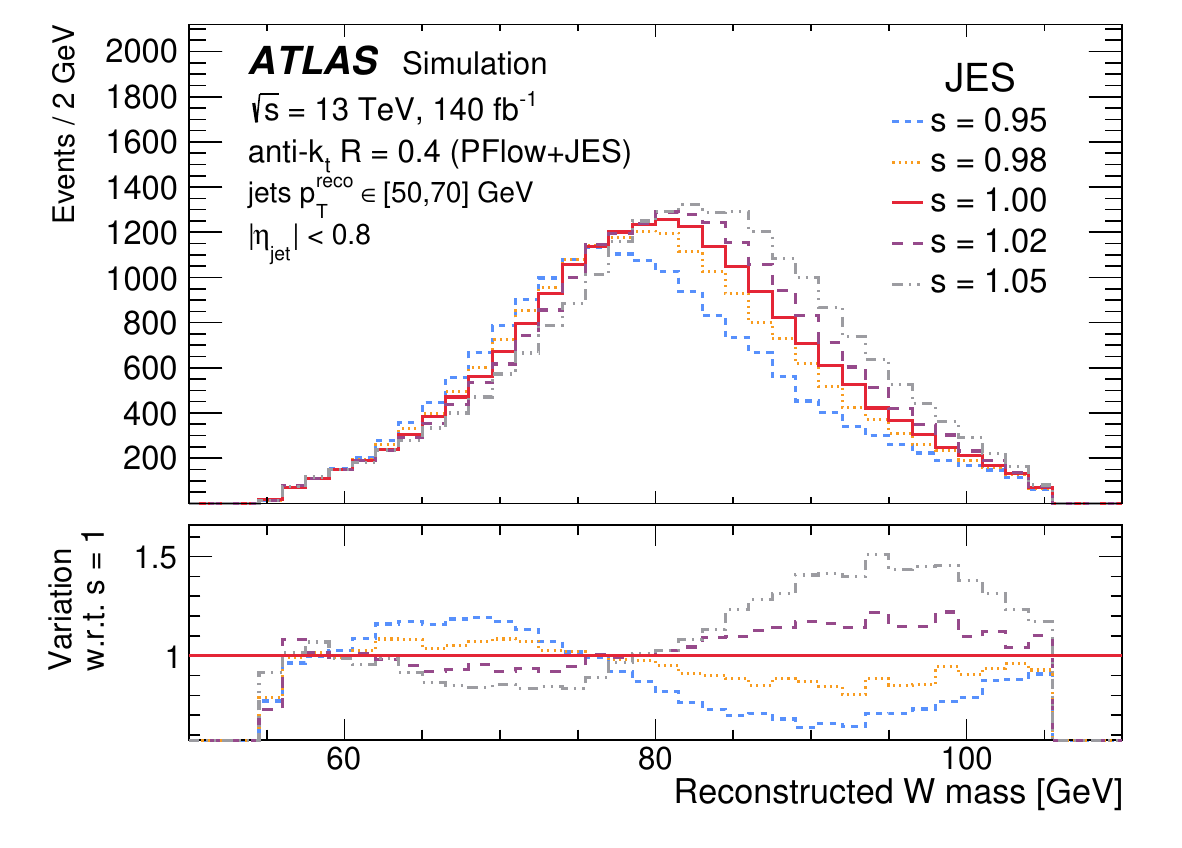}}\\
\subfloat[]{\includegraphics[width=0.80\textwidth]{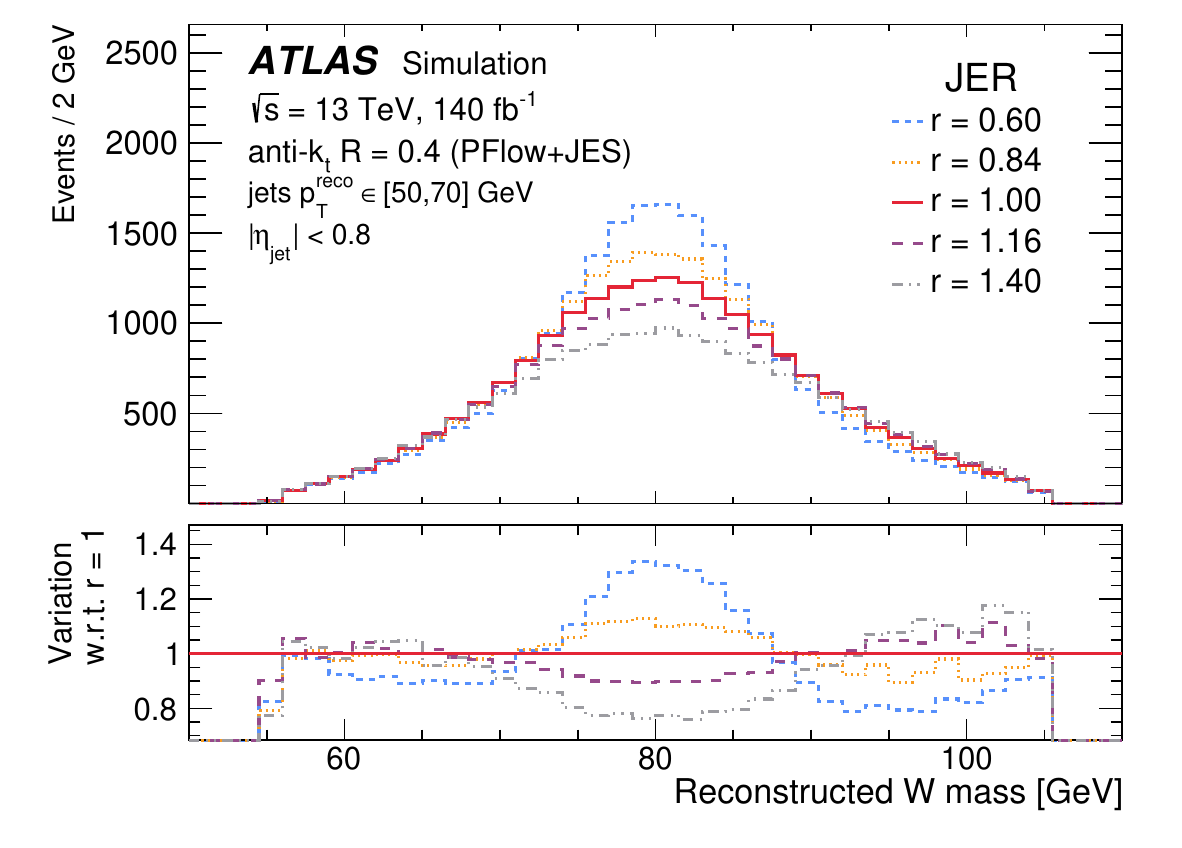}}
\caption{Reconstructed $W$-boson mass in the region with the two jets from the $W$-boson decay with reconstructed~\pt between 50 and 70~\gev in the Run~2 simulation. Templates with different (a) JES and (b) JER~assumptions represented by the parameters of the forward-folding formula are shown. For the JES (JER) variations shown, the JER (JES) correction is set to one. The differences between the total yields for each distribution come from the acceptance effects due to the changes in the reconstructed jet~\pt. The bottom panels show the ratios to the template with nominal (a) JES or (b) JER assumption. The solid line in the bottom panels represents the ratio of one.}
\label{fig:template_50_70}
\end{figure}

\FloatBarrier


%
\section{Systematic uncertainties}
\label{sec:systematics}

Several systematic uncertainties can affect the measurement of the JES and JER parameters.
Uncertainties affecting leptons, \met and $b$-tagged jets can impact the measurement due to the acceptance effects, while the uncertainties in the modelling can also affect the $W$-boson lineshape.

The uncertainties in the electron and muon identification, isolation, reconstruction, track-to-vertex association and trigger scale factors measured using $Z\to\mumu$ and $Z\to\ee$~\cite{EGAM-2018-01, MUON-2018-03, TRIG-2018-05, TRIG-2018-01} are evaluated.
Uncertainties related to the electron (muon) energy (momentum) scale and resolution are also evaluated as estimated by using resonance decays~\cite{EGAM-2018-01, MUON-2022-01}.

The uncertainty in the efficiency of the jet-vertex tagger algorithm is evaluated via the variation of the corrections applied to the simulation.
The JES and JER uncertainties from the best available calibration are estimated by using the Run~2 data sample~\cite{PERF-2014-03}.
The same uncertainties are applied to the Run~3 based simulation, and are validated using Run~3 data.
Additional uncertainties are considered for the Run~3 simulation to cover the residual differences between data and simulation.

The impact of the JER uncertainties is evaluated for the JES correction and the impact of the JES uncertainties is evaluated for the JER correction.
The uncertainties in the efficiency of $b$-tagging are evaluated via the variation of the scale factors affecting the $b$-/$c$-/light-flavour (in)efficiencies, using measurements similar to Refs.~\cite{FTAG-2018-01, FTAG-2019-02, FTAG-2020-08}.

The uncertainty from the leptons and jets is propagated to the \met.
Additionally, a dedicated set of uncertainties from the soft term is evaluated, varying the magnitude and the direction of the missing transverse momentum vector.

Several sources affecting the modelling of the \ttbar process are evaluated, affecting the production or decay of the top quarks.
Variations of the renormalisation and factorisation scale in the ME by factors of 0.5 and 2 are evaluated separately.
Uncertainty in the initial-state radiation (ISR) is estimated by using the \emph{Var3c} variation of the \PYTHIA8 A14 tune~\cite{ATL-PHYS-PUB-2014-021}.
Variations of the renormalisation scale in the strong coupling constant in the final-state radiation ($\alpha_{\text{s}}^{\text{FSR}}$) by factors of 0.5~and~2 of the A14 tune are evaluated.
The uncertainty related to the \POWHEG to \PYTHIA matching is estimated by comparing with a dedicated \POWPY[8] sample with the $\pt^{\text{hard}}$ parameter set to one instead of the default value of zero.
The uncertainty in the choice of the \hdamp parameter is estimated by comparing with a dedicated \POWPY[8] sample produced with twice the nominal \hdamp value (\hdamp = 3$m_{\text{t}}$).
Due to a known mismodelling of the top-quark~\pt distribution, an uncertainty estimated from reweighting the top-quark~\pt to NNLO QCD predictions~\cite{Czakon:2017wor} is evaluated as a proxy for missing higher order predictions in the kinematics.
This uncertainty is found to have a negligible impact on the JES or JER estimate in the Run~2 analysis and is thus not considered further.
The impact of using a different parton shower and hadronisation model on the Run~2~(Run~3) analysis is evaluated by comparing the nominal \ttbar sample with another sample generated with \POWHEG interfaced to \HERWIG~7.2.1~\cite{Bahr:2008pv, Bellm:2015jjp, Bellm:2017jjp} (\HERWIG~7.2.3), using the \HERWIG~7.2 default set of tuned parameters~\cite{Bellm:2019zci} and the \MMHT[lo] PDF set~\cite{Harland-Lang:2014zoa}.
The \POWHEG settings are the same as for the nominal sample.
To estimate the uncertainty in the colour reconnection (CR), samples were generated with three models: CR0 --- a multiple-parton-interactions (MPI) based model, CR1 --- a QCD-based model and CR2 --- a gluon-move based model \cite{ATL-PHYS-PUB-2017-008}.
The uncertainty is estimated by comparing the CR1 and CR2 models with the prediction of CR0 and taking the larger uncertainty.
Variations in the top-quark mass of 0.5~\gev are evaluated by using dedicated samples with $m_{\text{top}}=172.0~\gev$ and $m_{\text{top}}=173.0~\gev$.
Variation in the MPI and underlying event (UE) modelling (\emph{Var1} variation in the A14 tune) is estimated with samples with UE variations in \PYTHIA.
The modelling of secondary gluon emissions from quarks originating from coloured resonance decays can affect both the $b$-quark and $W$-boson kinematics as well as the $b$-jet fragmentation.
To assess the impact of this choice on the measurement, an alternative sample with the recoil model within \PYTHIA[8] changed to \emph{recoilToTop} was produced.
The uncertainty in the PDF estimate for the \ttbar process is assessed by evaluating the impact of the PDF4LHC~\cite{Butterworth:2015oua,Ball:2022oua} variations.
Since the $W$ bosons decay into an $s$-quark and a $c$-quark in about half of the hadronic decays, a variation of the Lund--Bowler $c$-fragmentation function~\cite{Andersson:1983jt,Bowler:1981sb} is tested, but it is found to have a negligible impact.

The uncertainty due to the choice of the parton shower generator for the $tW$ process is estimated by comparing the nominal \POWPY[8] sample with \POWHEG interfaced to the \HERWIG sample.
The variations in the ME, \emph{Var3c} and $\alpha_{\text{s}}^{\text{FSR}}$ are evaluated analogously to the \ttbar case.
The uncertainty in the overlap between the \ttbar and $tW$ diagrams is assessed by comparing the nominal prediction using the diagram-removal approach to the diagram-subtraction approach~\cite{Frixione:2008yi} in an alternative sample.

Uncertainties in the prediction of the cross-section of the various processes are taken into account although these are expected to have only a small impact on the measured $W$-boson line shape.
A 5\% uncertainty is considered for the \ttbar and single-top processes following the uncertainty in the predicted cross-section~\cite{Beneke:2011mq,Cacciari:2011hy,Baernreuther:2012ws,Czakon:2012zr,Czakon:2012pz,Czakon:2013goa,Czakon:2011xx,Kidonakis:2021vob}. A conservative 50\% uncertainty is assigned to the $V$+jets and $VV$+jets, and 100\% uncertainty is applied to the fake-lepton estimate.
These uncertainties in the predicted cross-section have a negligible impact on the measured corrections for the JES and JER.

\FloatBarrier


%
\section{Results}

The mean values $\mu$ of the $W$-boson mass distributions in the \pt regions defined in Section~\ref{sec:event_selection} are used to extract the JES parameters from a fit to data.
Similarly, the standard deviation $\sigma$ of the $W$-boson mass distribution is used for the JER measurement.
The $\mu$ ($\sigma$) of each forward-folded $W$-boson mass distribution is extracted for each value of the $s$ ($r$) parameters from the forward-folding Eq.~(\ref{eq:FF}) and in the case of the off-diagonal regions, for each combination of the $s_1$ and $s_2$ ($r_1$ and $r_2$) parameters of the JES (JER) estimate.
Due to the $\chi^2 < 1$ selection, the minimum and the maximum of the reconstructed $W$-boson mass distribution is bounded to the range $m_W^{\text{gen}} \pm 25$~\GeV, thus the $\sigma$ extraction does not suffer from outliers.
In the diagonal regions the obtained $\mu$ ($\sigma$) are fitted with a linear (quadratic) function $\mathcal{F}(s)$ ($\mathcal{F}(r)$).
In the off-diagonal regions, the $\mu$ ($\sigma$) are fitted with a quadratic function $\mathcal{F}(s_1,s_2)$ ($\mathcal{F}(r_1,r_2)$) where $s_1$ and $s_2$ ($r_1$ and $r_2$) are the forward-folding formula parameters for the leading and sub-leading jet identified to originate from the $W$-boson decay.
The diagonal parameterisation function $\mathcal{F}(s)$ and off-diagonal parameterisation function $\mathcal{F}(s_1,s_2)$ share a set of six parameters $s_{i}$, one for each reconstructed jet \pt bin, and analogously for the JER parameterisation.
For the JES parameterisation, most fitted points are within one standard deviation from the fitted function with only one point slightly above two standard deviation, with the standard deviation representing the uncertainty due to the finite number of simulated events.
For the JER parameterisation, most fitted points are within one standard deviation with one point around 1.5 standard deviations away from the fitted function.
Additionally, a dedicated uncertainty is assigned for the imperfect parameterisation and for neglecting possible correlations between the parameters in the $\mathcal{F}$ functions due to region migrations of simulated events when varying the JES and JER corrections and it is referred to as `non-closure' uncertainty.
Due to the low statistical power of the separation of the various JER smeared templates in the diagonal region with jet~\pt between 150 and 200~\gev, this region is not included in the JER correction measurement.
Figure~\ref{fig:Parametrisation_Run2} shows an example of the parameterisation for the JES and JER parameters in the Run~2 data.
The larger impact on $\sigma$ for the lower~\pt jet compared to the higher~\pt jet originates from the naturally lower resolution for the lower~\pt jets due to the detector's calorimeter response.

\begin{figure}[ht]
\centering
\subfloat[]{\includegraphics[width=0.49\textwidth]{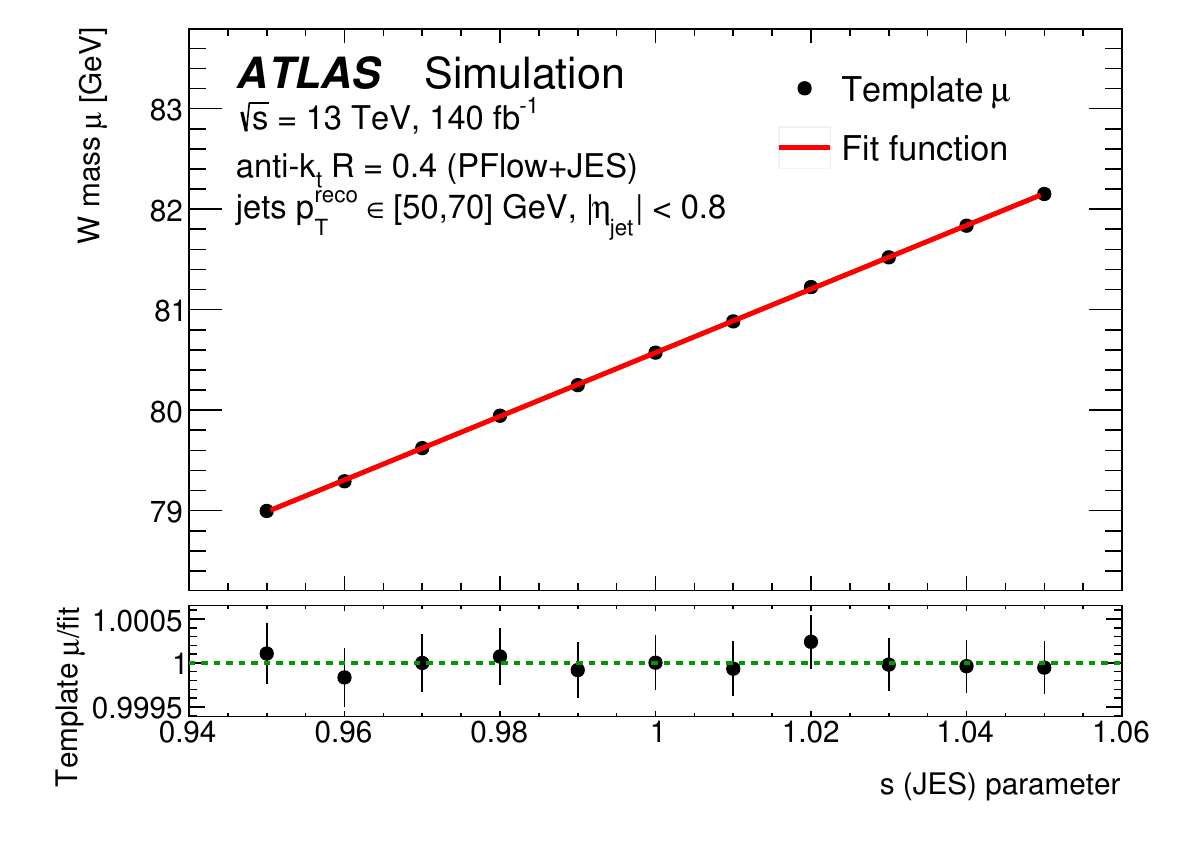}}
\subfloat[]{\includegraphics[width=0.49\textwidth]{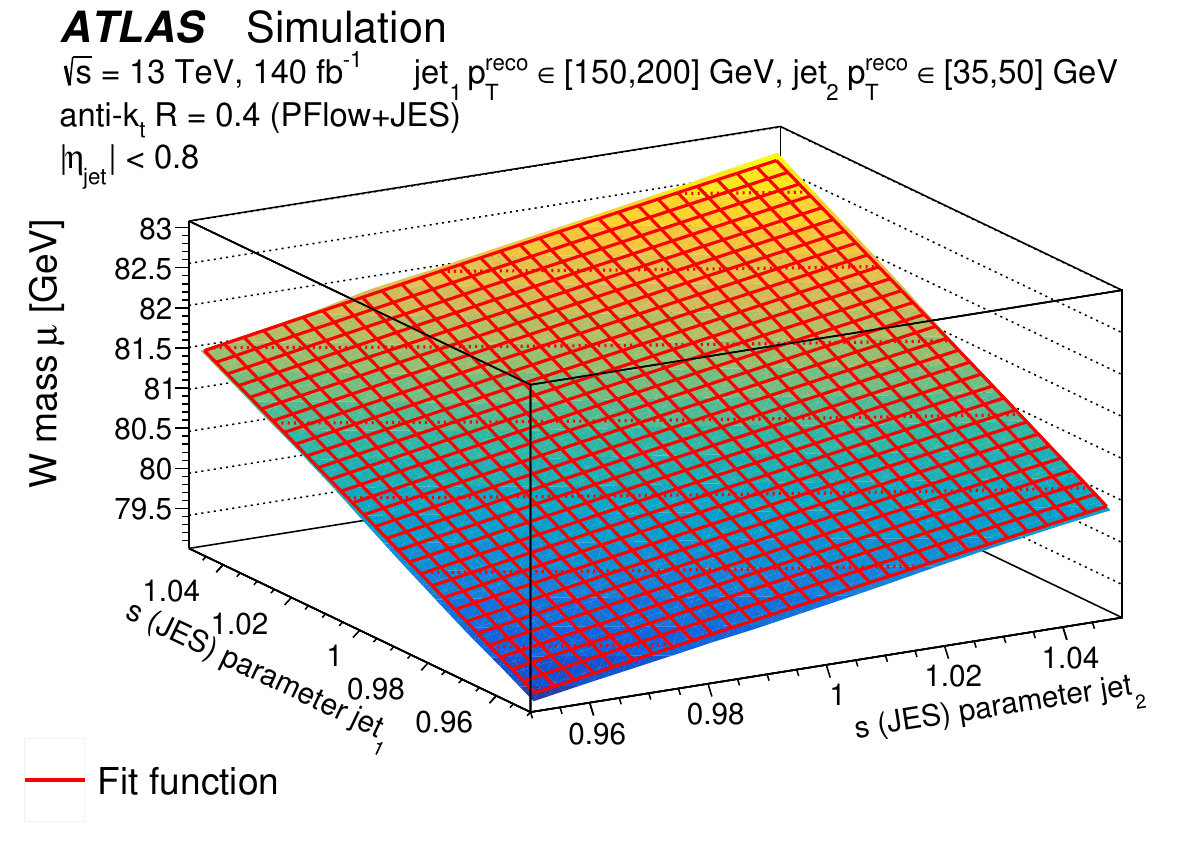}}\\
\subfloat[]{\includegraphics[width=0.49\textwidth]{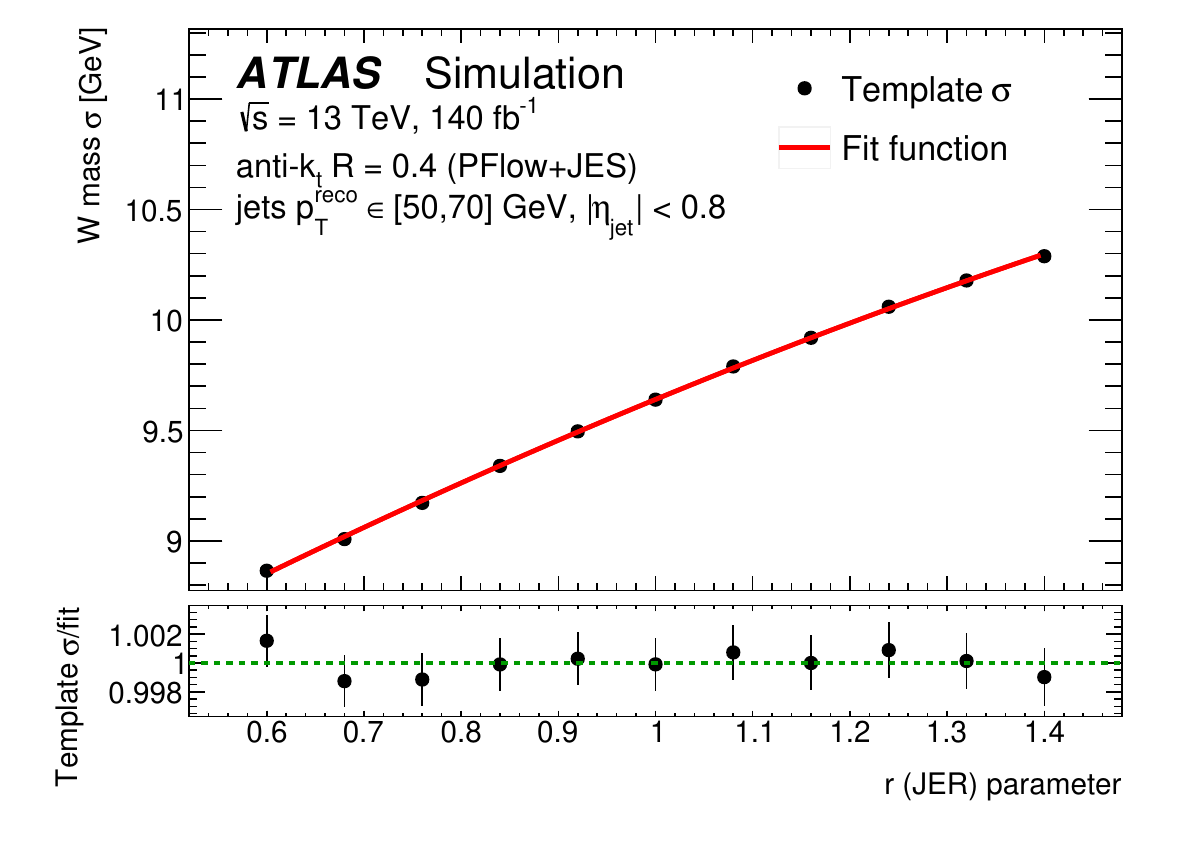}}
\subfloat[]{\includegraphics[width=0.49\textwidth]{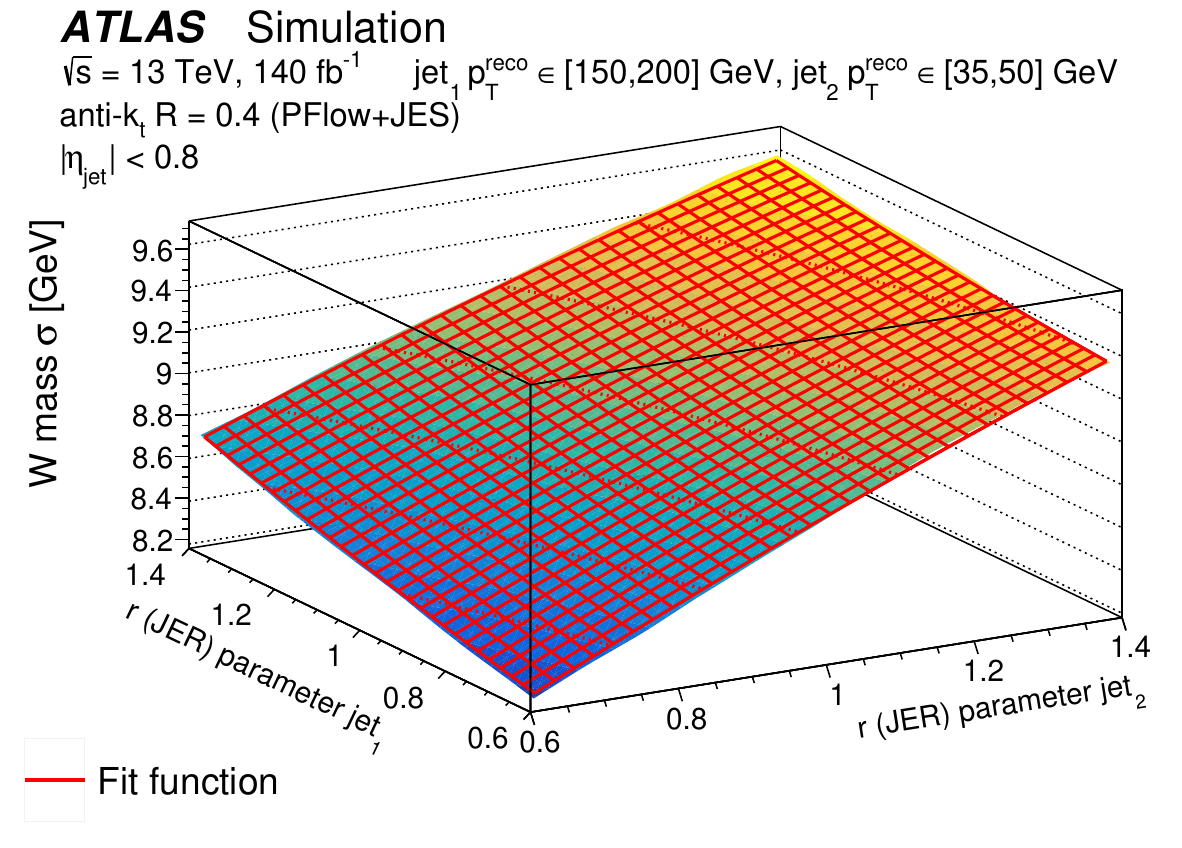}}
\caption{The mean (top plots) and standard deviation (bottom plots) of the $W$-boson mass distribution in a representative reconstructed jet~\pt region as a function of the JES~(JER) parameter in the Run~2 simulation. (a) and (c) show the diagonal region with reconstructed jet~\pt between 50 and 70~\gev, while (b) and (d) show the off-diagonal region with the leading jet ($\text{jet}_{\text{1}}$) with reconstructed~\pt between 150 and 200~\gev and the sub-leading jet ($\text{jet}_{\text{2}}$) with reconstructed~\pt between 35 and 50~\gev. The fitted function is shown by the solid line. For the diagonal regions, the bottom panels show the ratio of the mean~(standard deviation) value to the fitted function. The dashed line in the bottom panels represents the ratio of one. For the off-diagonal regions, the colour gradient shows the mean~(standard deviation) of the $W$-boson mass distribution for the JES~(JER) templates.}
\label{fig:Parametrisation_Run2}
\end{figure}

Once the parameterisations $\mathcal{F}$ are obtained, a maximum-likelihood fit to $\mu$ ($\sigma$) in data is performed to extract the values of the JES (JER) parameters for all the regions simultaneously.
The likelihood is defined as

\begin{equation*}
\mathcal{L}\left(\vec{\mu}|\vec{s}\right) = \prod_{i \in \text{reg}} G\left[\mu_i|\mathcal{F}_i(\vec{s}), \sigma_{\mu_i}\right],
\end{equation*}

where $\vec{\mu}$ is the vector of $\mu$ ($\sigma$) values measured in data, $\vec{s}$ is the vector of the scale (resolution) parameters to be estimated, $G$ represents the Gaussian function, $\mathcal{F}$ is the parameterisation of $\mu$ ($\sigma$) values as described above and $\sigma_{\mu_i}$ is the data statistical uncertainty in the $\mu$ ($\sigma$) values in region $i$ for the JES (JER) estimate.

The systematic uncertainties in the measurement of the JES and JER correction factors are estimated by shifting the expected distributions as summarised in Section~\ref{sec:systematics}, fitting the expected distributions with the nominal model and evaluating the impact on the best-fit parameters.
The parameterisation `non-closure' uncertainty is estimated by fitting the distributions representing a variation of up to 5\% in JES and up to 25\% in JER, and comparing the best-fit values with the $s$ and $r$ values used to generate the given varied distribution.
The final `non-closure' uncertainty in each \pt bin is estimated as the maximal deviation from the input value for all variations and is symmetrised.

Figures~\ref{fig:DataPostFit_JES} and~\ref{fig:DataPostFit_JER} show the $\mu$ or $\sigma$ of the $W$ mass distribution as obtained without forward-folding (pre-fit) and with forward-folding using the fitted JES or JER (post-fit) correction factors.
Because each region has different selections applied, the $\mu$ and $\sigma$ values vary accordingly, however the calibration procedure takes these differences into account.
Due to the non-Gaussian nature of the uncertainties, stemming from the non-linear parameterisation, pseudo-experiments are used to obtain the uncertainty band.
In the pseudo-experiments, random Gaussian distributed numbers are generated for the JES~(JER) parameters with the standard deviation representing the total measured uncertainty, and the variations are then propagated to $\mu$ ($\sigma$) of the $W$-boson mass distribution using the obtained parameterisation.
The diagonal region with jets between 150 and 200~\gev has a large statistical uncertainty due to the relatively low number of selected events compared to the other regions.

The fitted central values for the JES and JER parameters with their uncertainties are shown in Figure~\ref{fig:FinalResult} for both Run~2 and Run~3 data.
Bin-centre corrections are applied to the horizontal values by taking the mean of the jet~\pt distribution of the jets from the $W$-boson decay in each \pt bin.
Tables~\ref{tab:Systematics_Category_JES_Run2}--\ref{tab:Systematics_Category_JER_Run3} show the breakdown of the sources of systematic uncertainties for the JES and JER measurement for all jet~\pt bins, separately for the Run~2 and Run~3 data.
The central values for the JES measurement are below unity in the Run~2 data and in most jet~\pt bins in the Run~3 data, indicating that the energy scale in the simulation needs to be decreased to match the scale observed in data.
The measured JES correction factors in Run~2 are compatible with the previously measured corrections by ATLAS using the standard calibration techniques~\cite{JETM-2018-05}.
The measured JES correction factors in Run~2 are consistently about 2\% lower than the Run~3 correction factors, which are closer to unity, suggesting the MC simulation in Run~3 models the experimental data better than the Run~2 simulation.
The measured JER correction factors are consistent with unity in all jet~\pt bins.

Overall, the uncertainties are compatible between the Run~2 and Run~3 data.
For the JES measurement, the first jet~\pt bin suffers from large uncertainties due to low sensitivity because of the small separation in templates between the different JES correction factors for the lowest~\pt bin.
Apart from the first jet~\pt bin, the JES uncertainties range from about 0.93\% (0.98\%) to about 1.7\% (1.6\%) in Run~2 (Run~3).
The uncertainties in the JER measurement range from about 14\% (16\%) to about 22\% (28\%) in Run~2 (Run~3).
The dominant uncertainties affecting the JES measurement stem from the variation of the $\alpha_{\text{s}}^{\text{FSR}}$ in the \ttbar samples, the comparison between \PYTHIA and \HERWIG generators and the JER uncertainties.
A shift in the $\alpha_{\text{s}}^{\text{FSR}}$ parameter shifts the peak of the reconstructed $W$-boson mass distribution mimicking the effects of the JES variation, and the impact decreases with higher jet~\pt.

\begin{figure}[ht]
\centering
\subfloat[]{\includegraphics[width=0.95\textwidth]{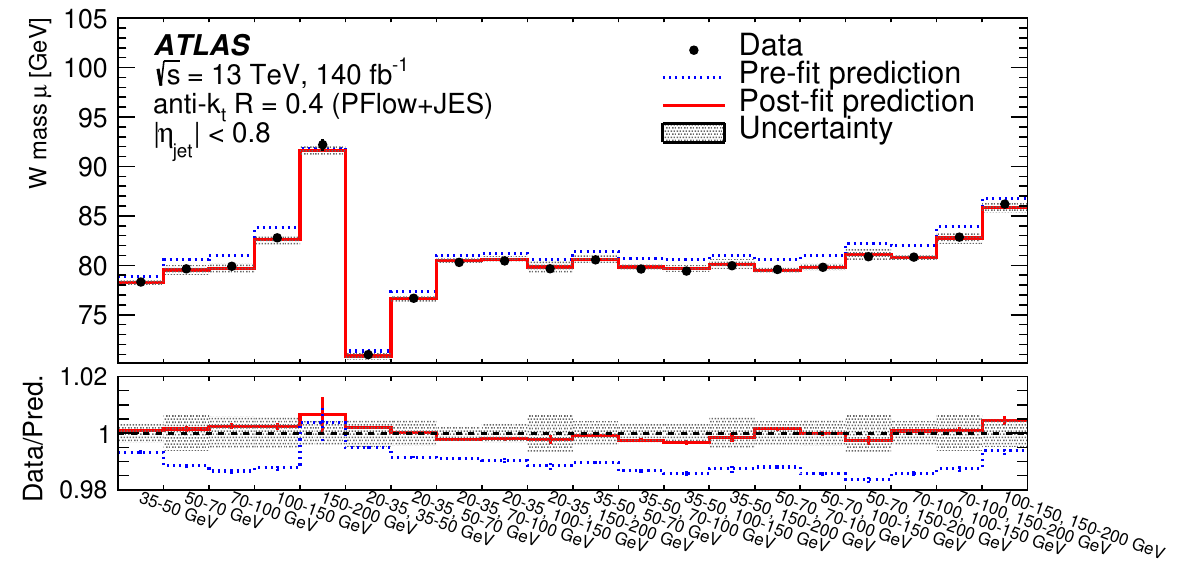}}\\
\subfloat[]{\includegraphics[width=0.95\textwidth]{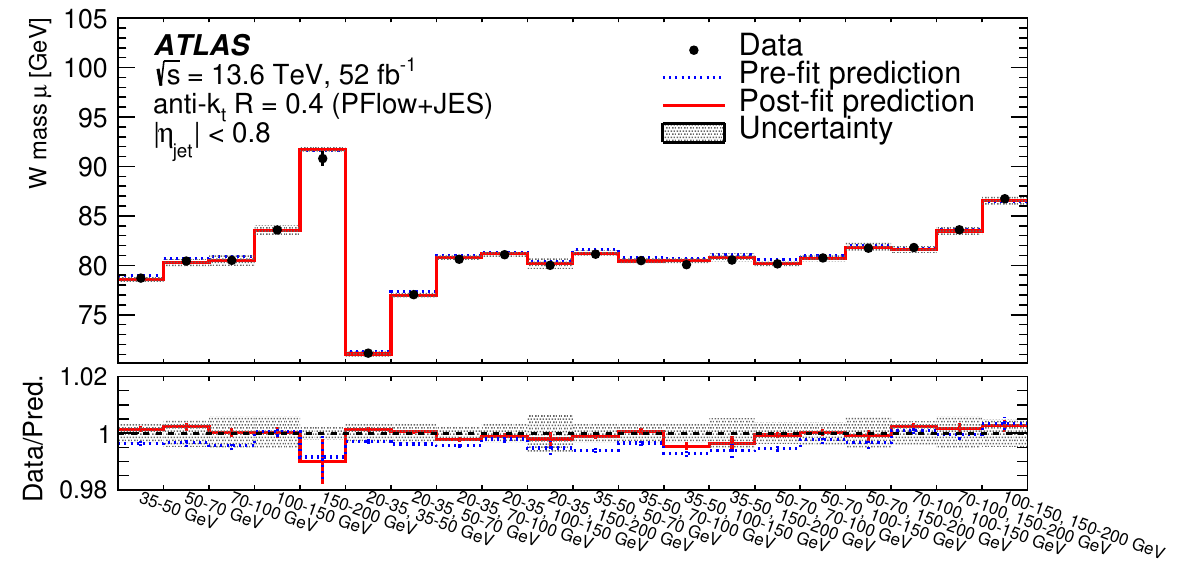}}
\caption{Mean of the $W$ mass distributions shown for the analysis regions defined in Section~\ref{sec:event_selection} in the JES fit in the (a) Run~2 and (b) Run~3 measurement. The values obtained in data are compared with those in simulations before and after the forward-folding correction. The uncertainties are obtained from pseudo-experiments taking into account the statistical and systematic uncertainties. The non-forward-folded (pre-fit) distribution is shown with a dotted line and the forward-folded (post-fit) distribution with a solid line. The first five bins show the diagonal regions, while the other bins show the off-diagonal regions. The bottom panels show the ratios of the data without the JES \textit{in situ} calibration (PFlow+JES) to the pre-fit (dotted line) and post-fit (solid line) predictions, with the dashed line representing the ratio of one.}
\label{fig:DataPostFit_JES}
\end{figure}

\begin{figure}[ht]
\centering
\subfloat[]{\includegraphics[width=0.95\textwidth]{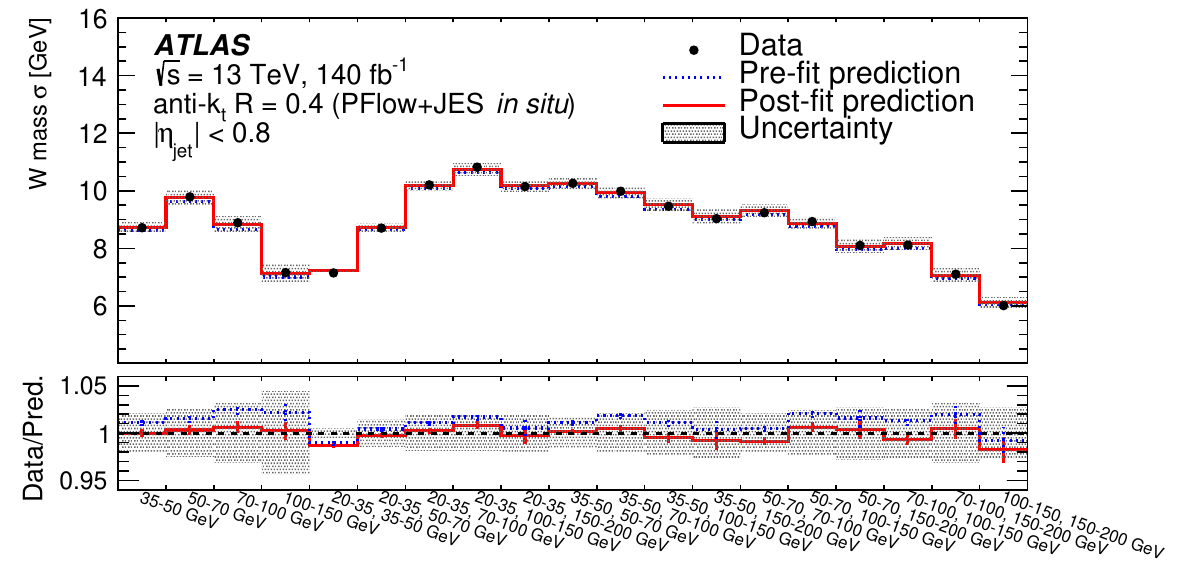}}\\
\subfloat[]{\includegraphics[width=0.95\textwidth]{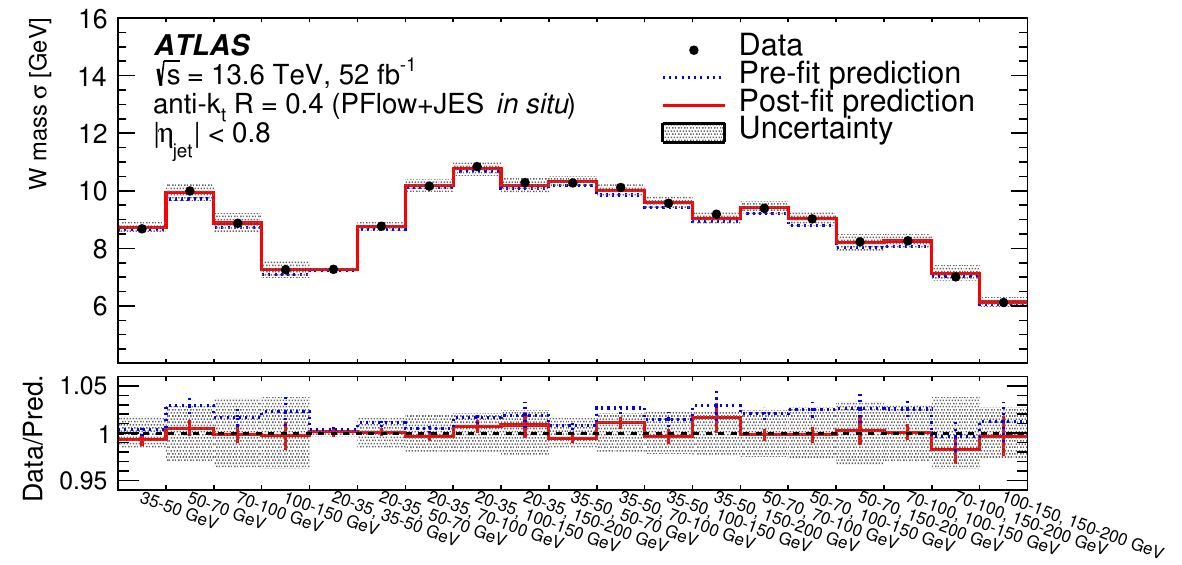}}
\caption{The standard deviation of the $W$ mass distributions shown for the analysis regions defined in Section~\ref{sec:event_selection} in the JER fit in the (a) Run~2 and (b) Run~3 measurement. The values obtained in data are compared with those in simulations before and after the forward-folding correction. The uncertainties are obtained from pseudo-experiments taking into account the statistical and systematic uncertainties. The non-forward-folded (pre-fit) distribution is shown with a dotted line and the forward-folded (post-fit) distribution with a solid line. The first four bins show the diagonal regions, while the other bins show the off-diagonal regions. The bottom panels show the ratios of the data with the JES \textit{in situ} calibration (PFlow+JES \textit{in situ}) to the pre-fit (dotted line) and post-fit (solid line) predictions, with the dashed line representing the ratio of one.}
\label{fig:DataPostFit_JER}
\end{figure}

\begin{figure}[ht]
\centering
\subfloat[]{\includegraphics[width=0.49\textwidth]{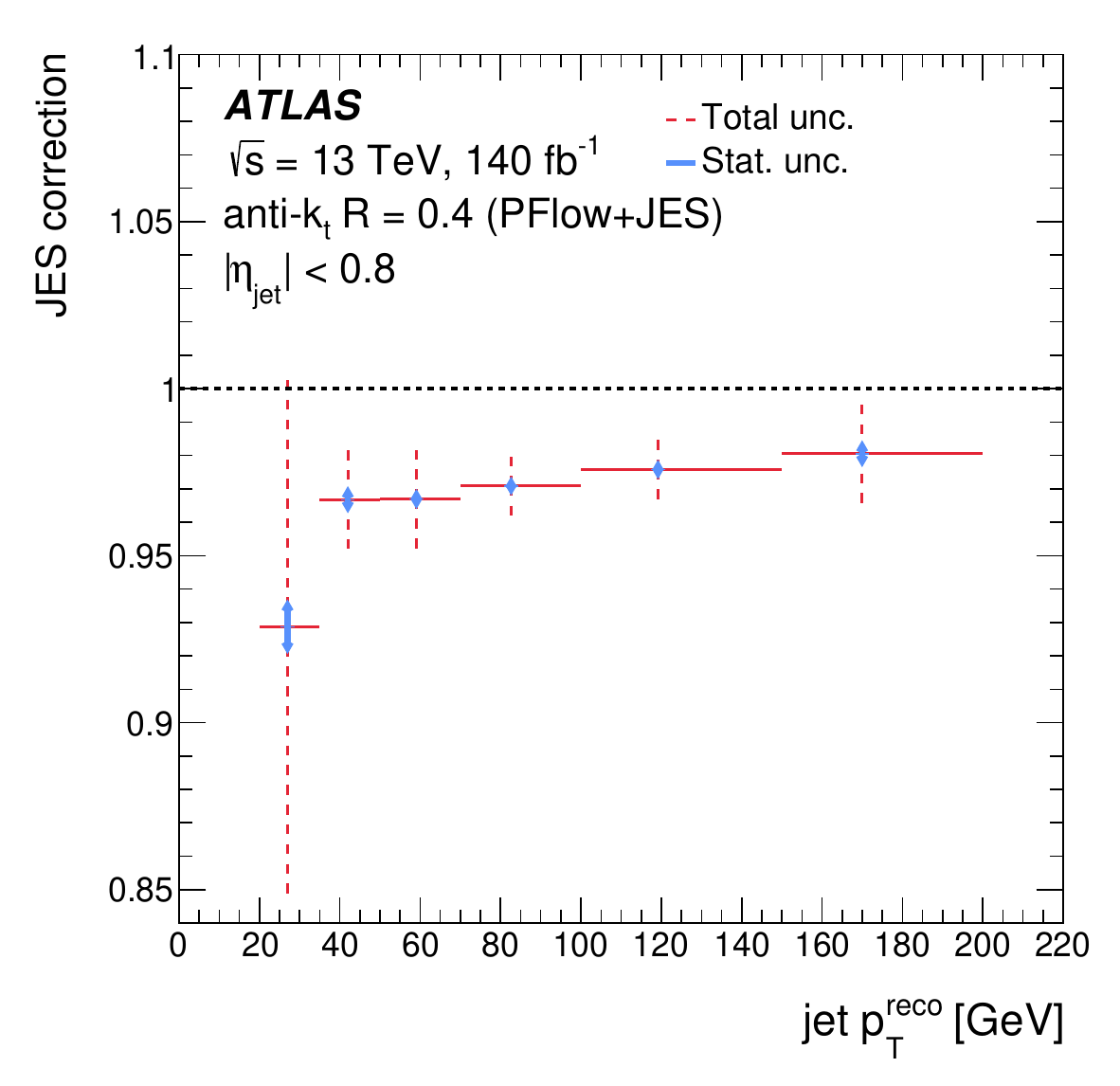}}
\subfloat[]{\includegraphics[width=0.49\textwidth]{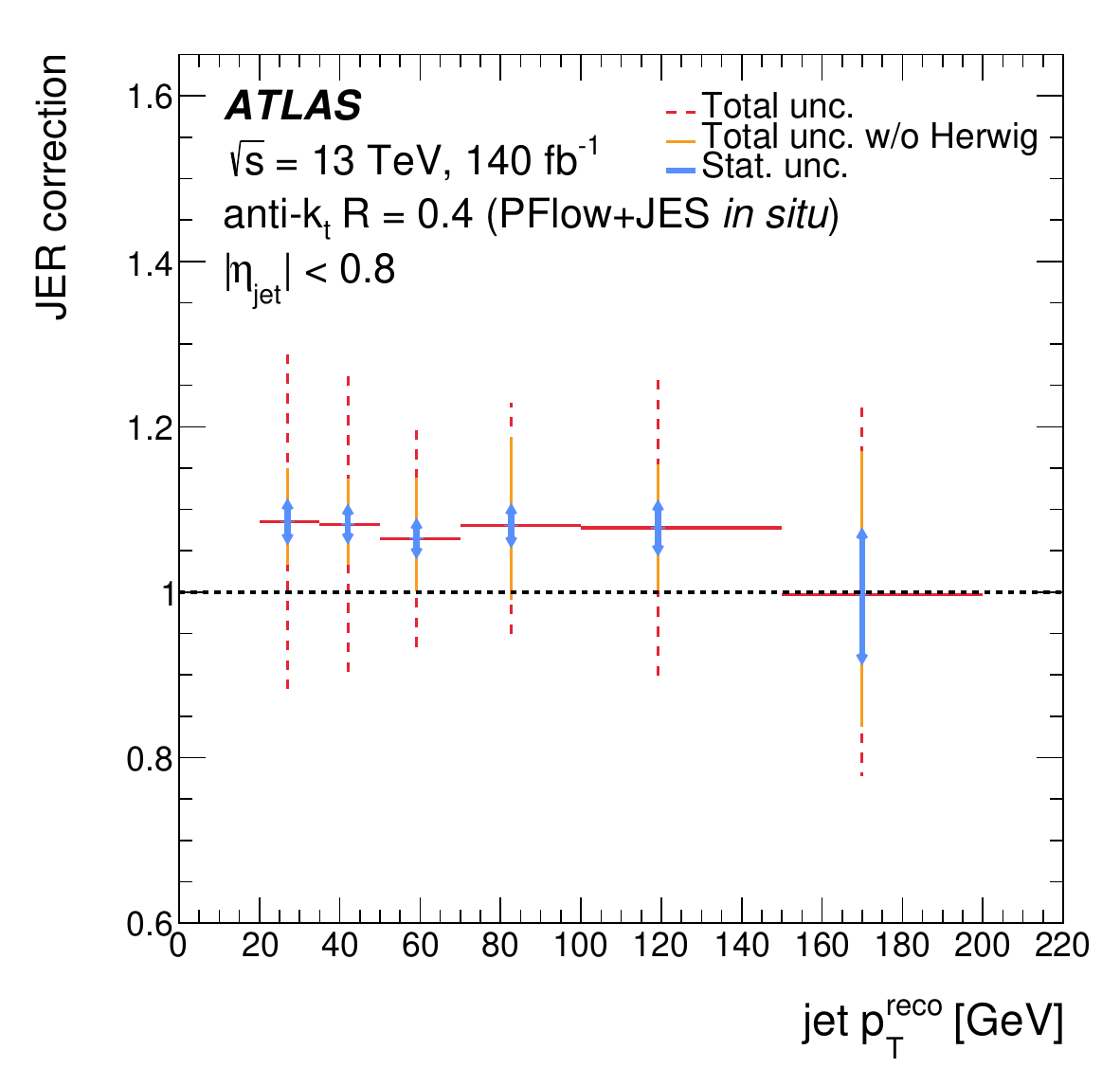}}\\
\subfloat[]{\includegraphics[width=0.49\textwidth]{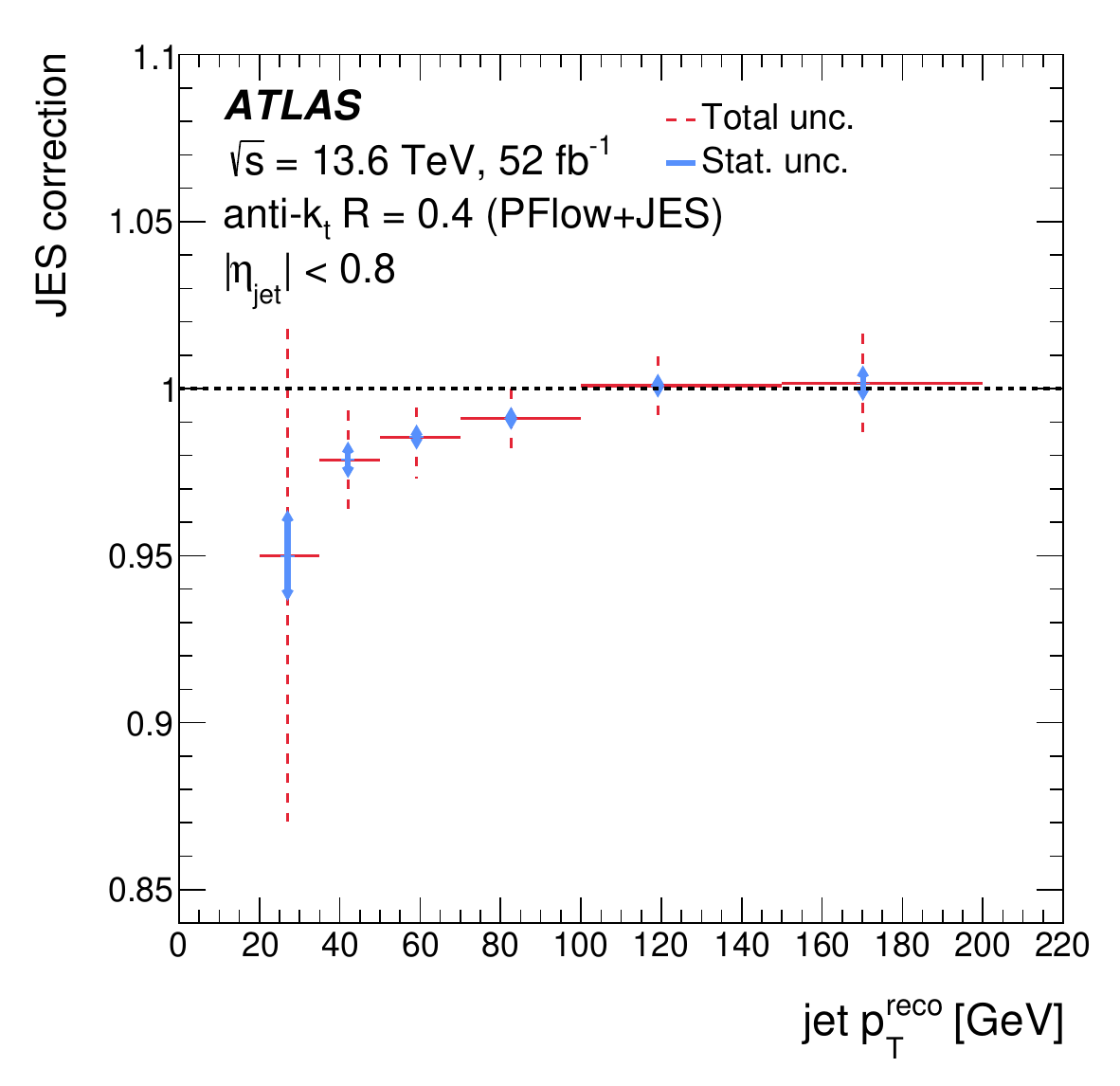}}
\subfloat[]{\includegraphics[width=0.49\textwidth]{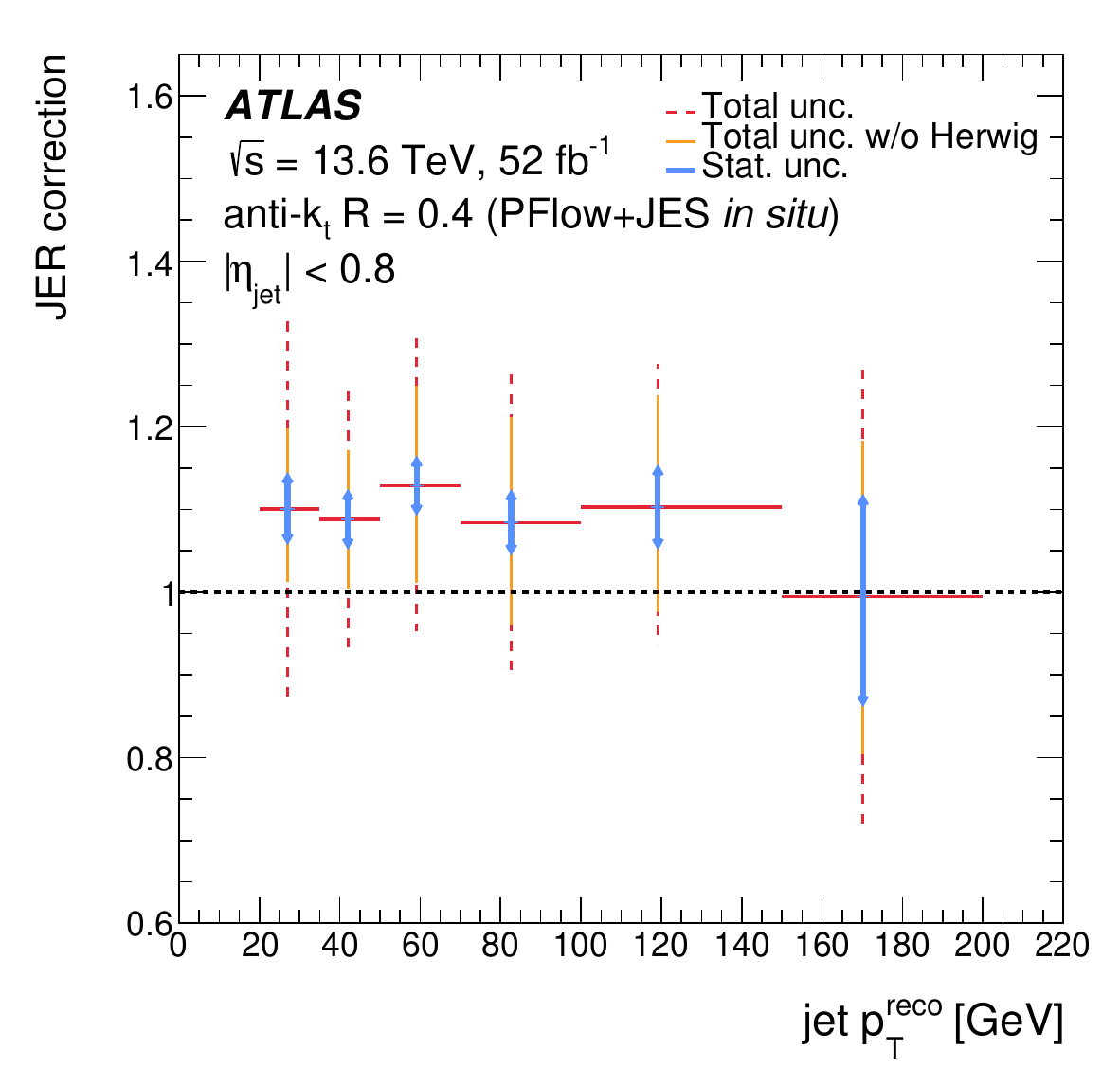}}
\caption{The central values and the corresponding uncertainties for the (a) Run~2 JES and (b) JER and (c) Run~3 JES and (d) JER correction factors as a function of reconstructed jet~\pt. The statistical uncertainty is shown with a thick solid bar with arrows on both ends, while the total uncertainty is shown with a thin dashed line. The thin solid line in the JER results (b, d) shows the total uncertainty without the \Pythia versus \Herwig component. The horizontal values are taken from the mean of the reconstructed jet~\pt distribution of the jets from the $W$-boson decay in each~\pt bin. The dashed horizontal line represents no correction.}
\label{fig:FinalResult}
\end{figure}

\FloatBarrier

\begin{table}[t]
\centering
\caption{Uncertainties in the JES estimate grouped into categories for the Run~2 data. Uncertainties for the different reconstructed jet~\pt regions are presented.}
\label{tab:Systematics_Category_JES_Run2}
\resizebox{\textwidth}{!}{%
\begin{tabular}{l|c|c|c|c|c|c}
\toprule
Category & \multicolumn{6}{c}{JES uncertainty [\%], Run~2} \\
\midrule
Region [\GeV]  & $20 < \pt^{\text{reco}} < 35$ & $35 < \pt^{\text{reco}} < 50$ & $50 < \pt^{\text{reco}} < 70$ & $70 < \pt^{\text{reco}} < 100$ & $100 < \pt^{\text{reco}} < 150$ & $150 < \pt^{\text{reco}} < 200$\\
\midrule
JER & 6.5\phantom{0} & 0.86 & 1.0\phantom{0} & 0.57 & 0.23 & 0.37\\
$b$-tagging & 0.09 & 0.04 & 0.03 & 0.02 & $<0.01$ & $<0.01$\\
Electron & 0.01 & $<0.01$ & $<0.01$ & $<0.01$ & $<0.01$ & $<0.01$\\
\met & 0.07 & 0.01 & $<0.01$ & $<0.01$ & $<0.01$ & 0.01\\
Muon & 0.02 & $<0.01$ & $<0.01$ & $<0.01$ & $<0.01$ & $<0.01$\\
\midrule
\ttbar \Pythia versus \Herwig & 2.8\phantom{0} & 0.63 & 1.0\phantom{0} & 0.71 & 0.83 & 1.6\phantom{0}\\
\ttbar $\alpha_{\text{s}}^{\text{FSR}}$ & 2.9\phantom{0} & 0.88 & 0.60 & 0.50 & 0.19 & 0.24\\
\ttbar CR variation & 0.98 & 0.21 & 0.05 & 0.17 & 0.15 & 0.10\\
\ttbar UE variation & 0.04 & 0.34 & 0.05 & 0.07 & 0.06 & 0.18\\
\ttbar other & 0.80 & 0.40 & 0.17 & 0.20 & 0.18 & 0.40\\
Single-top $tW$ modelling & 0.26 & 0.02 & 0.02 & 0.03 & 0.07 & 0.07\\
\midrule
Non-closure & 0.37 & 0.12 & 0.04 & 0.06 & 0.04 & 0.06\\
Normalisation & 0.09 & $<0.01$ & 0.02 & $<0.01$ & $<0.01$ & $<0.01$\\
Other & 0.13 & 0.03 & 0.04 & 0.06 & 0.07 & 0.09\\
\midrule
Statistical & 0.64 & 0.23 & 0.13 & 0.11 & 0.12 & 0.23\\
Total systematic & 7.8\phantom{0} & 1.5\phantom{0} & 1.6\phantom{0} & 1.1\phantom{0} & 0.92 & 1.7\phantom{0}\\
\midrule
Total uncertainty & 7.8\phantom{0} & 1.5\phantom{0} & 1.6\phantom{0} & 1.1\phantom{0} & 0.93 & 1.7\phantom{0}\\
\bottomrule
\end{tabular}
}
\end{table}

\begin{table}[b]
\centering
\caption{Uncertainties in the JER estimate grouped into categories for the Run~2 data. Uncertainties for the different reconstructed jet~\pt regions are presented.}
\resizebox{\textwidth}{!}{%
\begin{tabular}{l|c|c|c|c|c|c}
\toprule
Category & \multicolumn{6}{c}{JER uncertainty [\%], Run~2} \\
\midrule
Region [\GeV]  & $20 < \pt^{\text{reco}} < 35$ & $35 < \pt^{\text{reco}} < 50$ & $50 < \pt^{\text{reco}} < 70$ & $70 < \pt^{\text{reco}} < 100$ & $100 < \pt^{\text{reco}} < 150$ & $150 < \pt^{\text{reco}} < 200$\\
\midrule
JES & \phantom{0}3.1 & \phantom{0}1.6 & \phantom{0}5.8 & \phantom{0}8.2 & \phantom{0}6.1 & \phantom{0}6.1\\
$b$-tagging & \phantom{0}0.6 & \phantom{0}0.6 & \phantom{0}0.5 & \phantom{0}0.3 & \phantom{0}0.3 & \phantom{0}0.4\\
Electron & $<0.1$ & $<0.1$ & $<0.1$ & $<0.1$ & $<0.1$ & $<0.1$\\
\met & \phantom{0}0.1 & $<0.1$ & $<0.1$ & $<0.1$ & \phantom{0}0.1 & \phantom{0}0.3\\
Muon & $<0.1$ & $<0.1$ & $<0.1$ & $<0.1$ & $<0.1$ & \phantom{0}0.2\\
\midrule
\ttbar \Pythia versus \Herwig & 20\phantom{.0} & 17\phantom{.0} & 12\phantom{.0} & 10\phantom{.0} & 17\phantom{.0} & 15\phantom{.0}\\
\ttbar $\alpha_{\text{s}}^{\text{FSR}}$ & \phantom{0}3.3 & \phantom{0}3.1 & \phantom{0}1.6 & \phantom{0}1.6 & \phantom{0}1.3 & $<0.1$\\
\ttbar CR variation & \phantom{0}1.2 & \phantom{0}1.0 & \phantom{0}1.0 & \phantom{0}1.0 & \phantom{0}0.6 & 11\phantom{.0}\\
\ttbar UE variation & \phantom{0}0.8 & $<0.1$ & \phantom{0}0.7 & \phantom{0}1.4 & \phantom{0}1.1 & \phantom{0}5.9\\
\ttbar other & \phantom{0}1.8 & \phantom{0}2.4 & \phantom{0}2.2 & \phantom{0}4.3 & \phantom{0}3.1 & \phantom{0}4.7\\
Single-top $tW$ modelling & \phantom{0}0.2 & \phantom{0}0.2 & $<0.1$ & \phantom{0}0.3 & \phantom{0}0.2 & $<0.1$\\
\midrule
Non-closure & \phantom{0}0.6 & \phantom{0}0.6 & \phantom{0}0.7 & \phantom{0}0.4 & \phantom{0}0.6 & \phantom{0}1.6\\
Normalisation & \phantom{0}0.9 & \phantom{0}0.7 & \phantom{0}0.5 & \phantom{0}0.3 & \phantom{0}0.3 & \phantom{0}0.4\\
Other & \phantom{0}0.8 & \phantom{0}0.7 & \phantom{0}0.7 & \phantom{0}0.4 & \phantom{0}0.4 & \phantom{0}0.2\\
\midrule
Statistical & \phantom{0}2.1 & \phantom{0}1.8 & \phantom{0}1.8 & \phantom{0}2.0 & \phantom{0}2.8 & \phantom{0}7.6\\
Total systematic & 20\phantom{.0} & 18\phantom{.0} & 14\phantom{.0} & 14\phantom{.0} & 18\phantom{.0} & 21\phantom{.0}\\
\midrule
Total uncertainty & 20\phantom{.0} & 18\phantom{.0} & 14\phantom{.0} & 14\phantom{.0} & 19\phantom{.0} & 22\phantom{.0}\\
\bottomrule
\end{tabular}
}
\end{table}

\begin{table}[t]
\centering
\caption{Uncertainties in the JES estimate grouped into categories for the Run~3 data. Uncertainties for the different reconstructed jet~\pt regions are presented.}
\resizebox{\textwidth}{!}{%
\begin{tabular}{l|c|c|c|c|c|c}
\toprule
Category & \multicolumn{6}{c}{JES uncertainty [\%], Run~3} \\
\midrule
Region [\GeV]  & $20 < \pt^{\text{reco}} < 35$ & $35 < \pt^{\text{reco}} < 50$ & $50 < \pt^{\text{reco}} < 70$ & $70 < \pt^{\text{reco}} < 100$ & $100 < \pt^{\text{reco}} < 150$ & $150 < \pt^{\text{reco}} < 200$\\
\midrule
JER & 6.0\phantom{0} & 0.72 & 0.78 & 0.61 & 0.23 & 0.33\\
$b$-tagging & 0.10 & 0.05 & 0.04 & 0.03 & 0.02 & 0.01\\
Electron & 0.03 & $<0.01$ & $<0.01$ & $<0.01$ & $<0.01$ & $<0.01$\\
\met & 0.04 & 0.01 & 0.01 & $<0.01$ & $<0.01$ & 0.01\\
Muon & 0.03 & $<0.01$ & $<0.01$ & $<0.01$ & $<0.01$ & 0.01\\
\midrule
\ttbar \Pythia versus \Herwig & 2.0\phantom{0} & 0.20 & 0.64 & 0.71 & 0.83 & 1.4\phantom{0}\\
\ttbar $\alpha_{\text{s}}^{\text{FSR}}$ & 3.0\phantom{0} & 0.96 & 0.56 & 0.36 & 0.23 & $<0.01$\\
\ttbar CR variation & 0.40 & 0.38 & 0.09 & 0.04 & 0.09 & 0.12\\
\ttbar UE variation & 0.08 & 0.46 & 0.27 & 0.24 & 0.16 & 0.41\\
\ttbar other & 1.7\phantom{0} & 0.59 & 0.09 & 0.17 & 0.33 & 0.48\\
Single-top $tW$ modelling & 0.22 & $<0.01$ & 0.01 & 0.03 & $<0.01$ & 0.01\\
\midrule
Non-closure & 0.43 & 0.18 & 0.07 & 0.03 & 0.06 & 0.10\\
Normalisation & 0.14 & 0.03 & 0.02 & $<0.01$ & $<0.01$ & $<0.01$\\
Other & 0.21 & 0.02 & 0.05 & 0.04 & 0.05 & 0.07\\
\midrule
Statistical & 1.2\phantom{0} & 0.37 & 0.20 & 0.17 & 0.18 & 0.36\\
Total systematic & 7.3\phantom{0} & 1.5\phantom{0} & 1.2\phantom{0} & 1.1\phantom{0} & 0.97 & 1.6\phantom{0}\\
\midrule
Total uncertainty & 7.4\phantom{0} & 1.5\phantom{0} & 1.2\phantom{0} & 1.1\phantom{0} & 0.98 & 1.6\phantom{0}\\
\bottomrule
\end{tabular}
}
\end{table}

\begin{table}[b]
\centering
\caption{Uncertainties in the JER estimate grouped into categories for the Run~3 data. Uncertainties for the different reconstructed jet~\pt regions are presented.}
\label{tab:Systematics_Category_JER_Run3}
\resizebox{\textwidth}{!}{%
\begin{tabular}{l|c|c|c|c|c|c}
\toprule
Category & \multicolumn{6}{c}{JER uncertainty [\%], Run~3} \\
\midrule
Region [\GeV]  & $20 < \pt^{\text{reco}} < 35$ & $35 < \pt^{\text{reco}} < 50$ & $50 < \pt^{\text{reco}} < 70$ & $70 < \pt^{\text{reco}} < 100$ & $100 < \pt^{\text{reco}} < 150$ & $150 < \pt^{\text{reco}} < 200$\\
\midrule
JES & \phantom{0}3.3 & \phantom{0}5.0 & \phantom{0}9.5 & 11\phantom{.0} & 11\phantom{.0} & 11\phantom{.0}\\
$b$-tagging & \phantom{0}0.7 & \phantom{0}0.7 & \phantom{0}0.6 & \phantom{0}0.3 & \phantom{0}0.4 & \phantom{0}0.8\\
Electron & $<0.1$ & $<0.1$ & $<0.1$ & $<0.1$ & $<0.1$ & \phantom{0}0.1\\
\met & \phantom{0}0.1 & \phantom{0}0.3 & \phantom{0}0.2 & $<0.1$ & \phantom{0}0.2 & \phantom{0}0.4\\
Muon & $<0.1$ & $<0.1$ & $<0.1$ & $<0.1$ & $<0.1$ & \phantom{0}0.1\\
\midrule
\ttbar \Pythia versus \Herwig & 22\phantom{.0} & 14\phantom{.0} & 13\phantom{.0} & 13\phantom{.0} & 11\phantom{.0} & 21\phantom{.0}\\
\ttbar $\alpha_{\text{s}}^{\text{FSR}}$ & \phantom{0}4.5 & \phantom{0}4.0 & \phantom{0}2.7 & \phantom{0}1.2 & \phantom{0}1.3 & \phantom{0}1.7\\
\ttbar CR variation & \phantom{0}2.6 & \phantom{0}2.5 & \phantom{0}3.8 & \phantom{0}2.3 & \phantom{0}3.6 & \phantom{0}7.0\\
\ttbar UE variation & \phantom{0}4.1 & \phantom{0}1.0 & \phantom{0}0.1 & \phantom{0}1.1 & \phantom{0}2.7 & \phantom{0}0.4\\
\ttbar other & \phantom{0}3.8 & \phantom{0}2.6 & \phantom{0}4.1 & \phantom{0}5.2 & \phantom{0}3.8 & \phantom{0}4.9\\
Single-top $tW$ modelling & \phantom{0}0.2 & \phantom{0}0.2 & \phantom{0}0.3 & \phantom{0}0.1 & \phantom{0}0.3 & \phantom{0}0.9\\
\midrule
Non-closure & \phantom{0}1.1 & \phantom{0}0.7 & \phantom{0}0.8 & \phantom{0}0.3 & \phantom{0}1.0 & \phantom{0}2.2\\
Normalisation & \phantom{0}0.7 & \phantom{0}0.6 & \phantom{0}0.6 & \phantom{0}0.2 & \phantom{0}0.3 & \phantom{0}0.4\\
Other & \phantom{0}0.9 & \phantom{0}0.7 & \phantom{0}0.8 & \phantom{0}0.5 & \phantom{0}0.5 & \phantom{0}1.4\\
\midrule
Statistical & \phantom{0}3.7 & \phantom{0}3.0 & \phantom{0}3.0 & \phantom{0}3.3 & \phantom{0}4.5 & 12\phantom{.0}\\
Total systematic & 23\phantom{.0} & 16\phantom{.0} & 17\phantom{.0} & 18\phantom{.0} & 16\phantom{.0} & 25\phantom{.0}\\
\midrule
Total uncertainty & 24\phantom{.0} & 16\phantom{.0} & 18\phantom{.0} & 18\phantom{.0} & 17\phantom{.0} & 28\phantom{.0}\\
\bottomrule
\end{tabular}
}
\end{table}

\FloatBarrier

The JER uncertainties are dominated by the difference between \PYTHIA and \HERWIG that accounts for most of the total uncertainty.
The difference can be traced to the large difference between the predictions of the $W$-boson lineshape in the generator output before the detector simulation and only part of the difference comes from the difference between the jet responses.
The difference between the generators resembles resolution effects, thus the difference is highly correlated with the JER effects, resulting in a large uncertainty from the comparison of the two parton-shower generators.

Figure~\ref{fig:DataPostFit_Wmass} compares the pre-fit and post-fit predictions and the observed data in the $W$-boson mass distribution for jets between 70 and 100~\gev in the Run~2 data.
The post-fit prediction is obtained by propagating the fitted JES values to the predicted yield in each bin.
The post-fit prediction agrees with data for all jet~\pt regions.

\begin{figure}[h]
\centering
\subfloat[]{\includegraphics[width=0.49\textwidth]{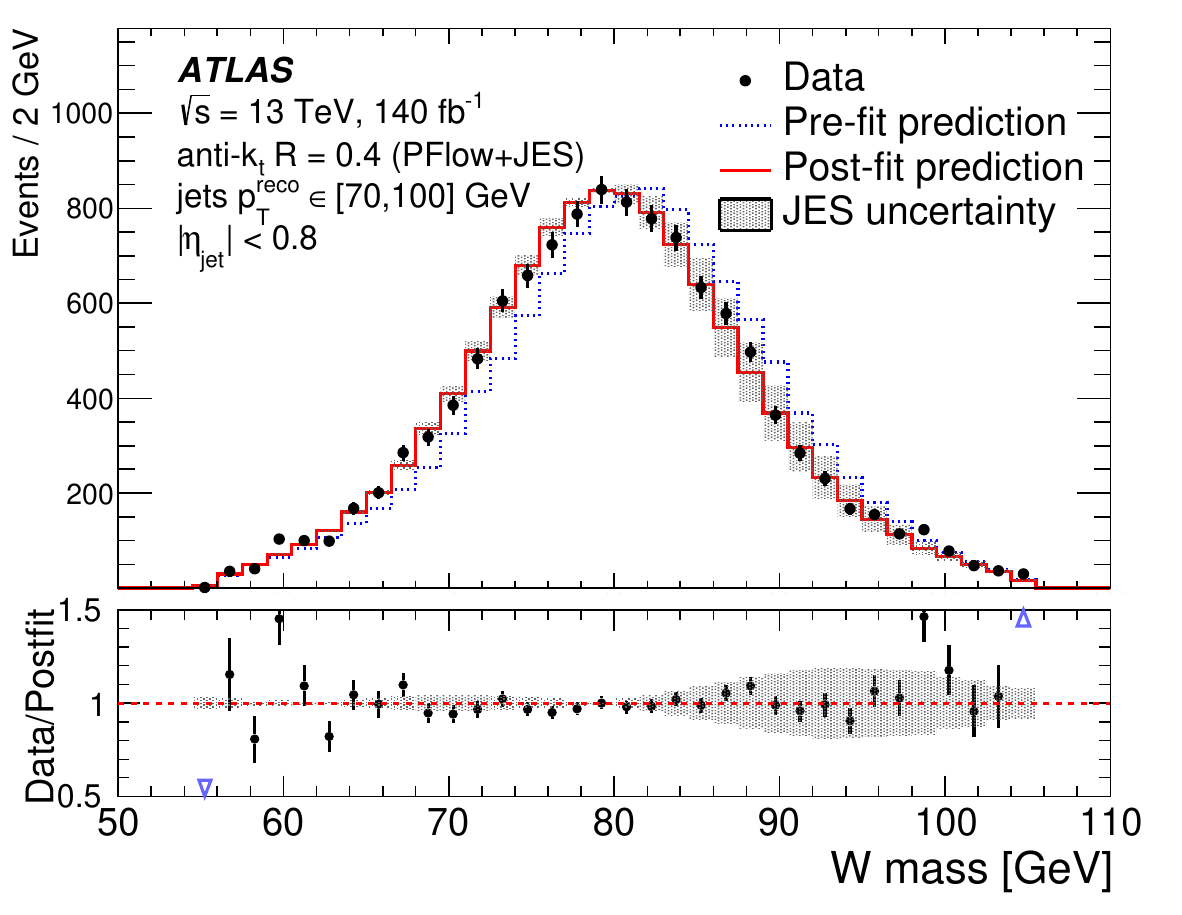}}
\subfloat[]{\includegraphics[width=0.49\textwidth]{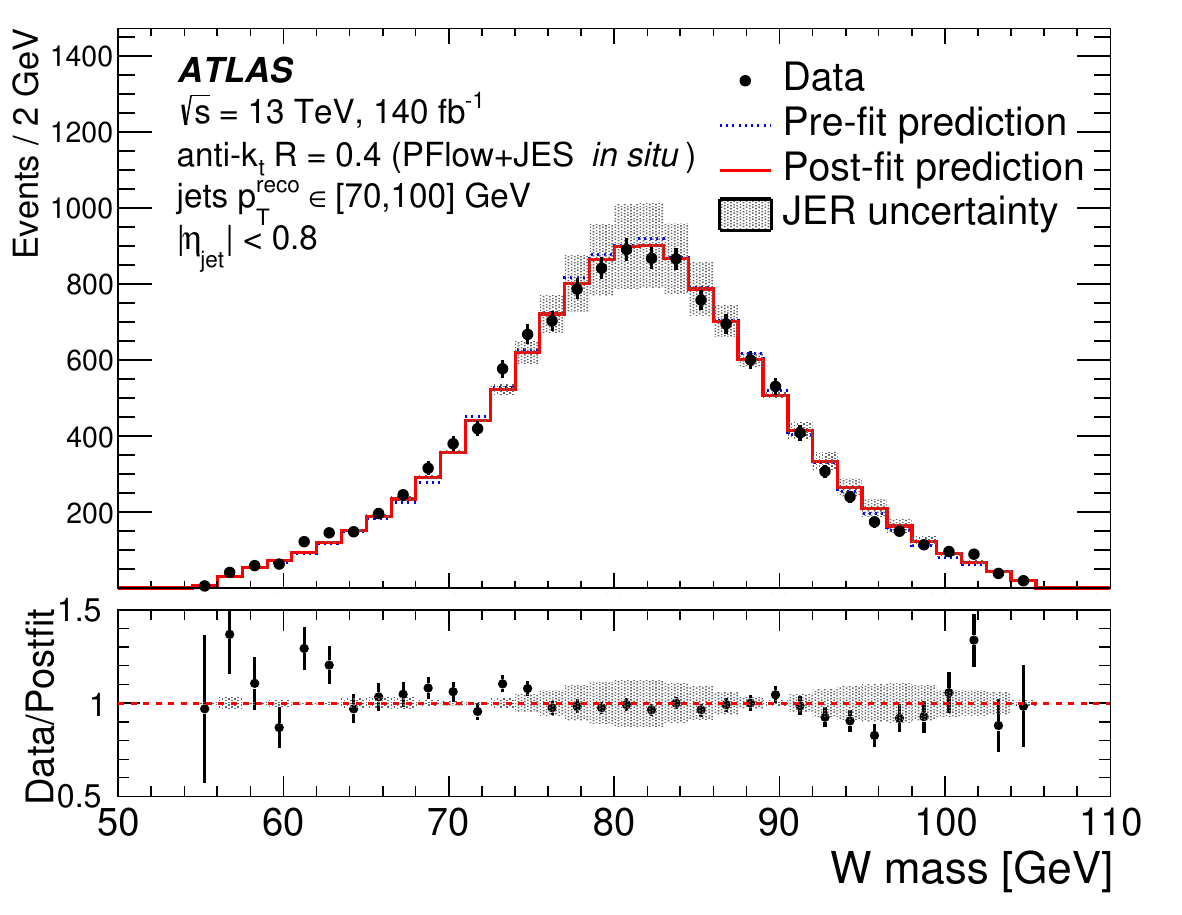}}
\caption{The pre-fit (dotted line) and post-fit (solid line) predictions for the $W$-boson mass distribution in the diagonal region with reconstructed jet~\pt between 70 and 100~\gev in the (a) JES and (b) JER fit to the Run~2 data. The hatched uncertainty band is estimated by propagating the total uncertainty in the (a) JES or (b) JER to the bin yield predictions. The predictions are normalised to the same total yield as data. The bottom panels show the ratio of the data (a) without the JES \textit{in situ} calibration (PFlow+JES) or (b) with the JES \textit{in situ} calibration (PFlow+JES \textit{in situ}) to the post-fit prediction, with the dashed horizontal line representing the ratio of one. The arrows indicate a data value outside of the displayed range.}
\label{fig:DataPostFit_Wmass}
\end{figure}

\FloatBarrier


%
\section{Conclusion}

A data-based calibration of the small-radius jet energy scale and resolution using \ttbar events with the ATLAS detector is presented.
The forward-folding technique is used to generate templates for the invariant mass of hadronically decaying $W$ bosons that are then compared with the distribution in data.
Two separate measurements are made, one for the full Run~2 data sample of the LHC at $\sqs = 13~\tev$ corresponding to an integrated luminosity of $140~\ifb$ and the other for the LHC Run~3 data sample collected in years 2022 and 2023, corresponding to an integrated luminosity of $52~\ifb$.
The extracted values for correction of the energy scale and resolution are binned in the jet transverse momentum ranging from 20 to 200~\gev.
This technique provides a precision that is competitive with the established calibration techniques for jet~$\pt > 35~\gev$.
However, analyses with a similar final state, especially those using observables correlated with the $W$ mass distribution as presented here, such as top-quark mass measurements, need to carefully evaluate whether this calibration can be applied to their measurements.
The dominant uncertainties for the JES and JER calibration originate from the modelling of the final-state radiation and parton shower in the \ttbar process.
If the dominant \ttbar modelling uncertainties are reduced in the future, a precision of about 0.5\% for the jet energy scale and about 10\% for the jet energy resolution could be reached using this \ttbar calibration technique on its own.
The results will be combined with the standard ATLAS calibration techniques.


%
\section*{Acknowledgements}
%

%

%
%

%
%

We thank CERN for the very successful operation of the LHC and its injectors, as well as the support staff at
CERN and at our institutions worldwide without whom ATLAS could not be operated efficiently.

The crucial computing support from all WLCG partners is acknowledged gratefully, in particular from CERN, the ATLAS Tier-1 facilities at TRIUMF/SFU (Canada), NDGF (Denmark, Norway, Sweden), CC-IN2P3 (France), KIT/GridKA (Germany), INFN-CNAF (Italy), NL-T1 (Netherlands), PIC (Spain), RAL (UK) and BNL (USA), the Tier-2 facilities worldwide and large non-WLCG resource providers. Major contributors of computing resources are listed in Ref.~\cite{ATL-SOFT-PUB-2026-001}.

We gratefully acknowledge the support of ANPCyT, Argentina; YerPhI, Armenia; ARC, Australia; BMWFW and FWF, Austria; ANAS, Azerbaijan; CNPq and FAPESP, Brazil; NSERC, NRC and CFI, Canada; CERN; ANID, Chile; CAS, MOST and NSFC, China; Minciencias, Colombia; MEYS CR, Czech Republic; DNRF and DNSRC, Denmark; IN2P3-CNRS and CEA-DRF/IRFU, France; SRNSFG, Georgia; BMFTR, HGF and MPG, Germany; GSRI, Greece; RGC and Hong Kong SAR, China; ICHEP and Academy of Sciences and Humanities, Israel; INFN, Italy; MEXT and JSPS, Japan; CNRST, Morocco; NWO, Netherlands; RCN, Norway; MNiSW, Poland; FCT, Portugal; MNE/IFA, Romania; MSTDI, Serbia; MSSR, Slovakia; ARIS and MVZI, Slovenia; DSI/NRF, South Africa; MICIU/AEI, Spain; SRC and Wallenberg Foundation, Sweden; SERI, SNSF and Cantons of Bern and Geneva, Switzerland; NSTC, Taipei; TENMAK, T\"urkiye; STFC/UKRI, United Kingdom; DOE and NSF, United States of America.

Individual groups and members have received support from BCKDF, CANARIE, CRC and DRAC, Canada; CERN-CZ, FORTE and PRIMUS, Czech Republic; COST, ERC, ERDF, Horizon 2020 and Marie Sk{\l}odowska-Curie Actions, European Union; Investissements d'Avenir Labex, Investissements d'Avenir Idex and ANR, France; DFG and AvH Foundation, Germany; Herakleitos, Thales and Aristeia programmes co-financed by EU-ESF and the Greek NSRF, Greece; BSF-NSF and MINERVA, Israel; NCN and NAWA, Poland; La Caixa Banking Foundation, CERCA Programme Generalitat de Catalunya and PROMETEO and GenT Programmes Generalitat Valenciana, Spain; G\"{o}ran Gustafssons Stiftelse, Sweden; The Royal Society and Leverhulme Trust, United Kingdom; United States of America.

In addition, individual members wish to acknowledge support from Chile: Agencia Nacional de Investigaci\'on y Desarrollo (ANID FONDECYT reg. 1230987, FONDECYT 1230812, FONDECYT 1240864, Fondecyt 3240661, Fondecyt Regular 1240721); China: Chinese Ministry of Science and Technology (MOST-2023YFA1605700, MOST-2023YFA1609300), National Natural Science Foundation of China (NSFC - 12175119, NSFC 12275265); Czech Republic: Czech Science Foundation (GACR - 24-11373S), Ministry of Education Youth and Sports (ERC-CZ-LL2327, FORTE CZ.02.01.01/00/22\_008/0004632), PRIMUS Research Programme (PRIMUS/21/SCI/017); EU: H2020 European Research Council (ERC - 101002463); European Union: European Research Council (BARD No. 101116429, ERC - 948254, ERC 101089007), European Regional Development Fund (HE COFUND GA No.101081355, ERDF), European Union, Future Artificial Intelligence Research (FAIR-NextGenerationEU PE00000013), Marie Sklodowska-Curie Actions (GAP-101168829); France: Agence Nationale de la Recherche (ANR-21-CE31-0013, ANR-21-CE31-0022, ANR-22-EDIR-0002, ANR-24-CE31-0504-01); Germany: Deutsche Forschungsgemeinschaft (DFG - 469666862, DFG - CR 312/5-2); China: Research Grants Council (GRF); Italy: Ministero dell'Università e della Ricerca (NextGenEU 153D23001490006 M4C2.1.1, NextGenEU I53D23000820006 M4C2.1.1, NextGenEU I53D23001490006 M4C2.1.1, SOE2024\_0000023); Japan: Japan Society for the Promotion of Science (JSPS KAKENHI JP22H01227, JSPS KAKENHI JP22H04944, JSPS KAKENHI JP22KK0227, JSPS KAKENHI JP24K23939, JSPS KAKENHI JP24KK0251, JSPS KAKENHI JP25H00650, JSPS KAKENHI JP25H01291, JSPS KAKENHI JP25K01023); Norway: Research Council of Norway (RCN-314472); Poland: Ministry of Science and Higher Education (IDUB AGH, POB8, D4 no 9722), Polish National Science Centre (NCN 2021/42/E/ST2/00350, NCN OPUS 2023/51/B/ST2/02507, NCN OPUS nr 2022/47/B/ST2/03059, NCN UMO-2019/34/E/ST2/00393, UMO-2022/47/O/ST2/00148, UMO-2023/49/B/ST2/04085, UMO-2023/51/B/ST2/00920, UMO-2024/53/N/ST2/00869); Portugal: Foundation for Science and Technology (FCT); Spain: Agencia de Gestión de Ayudas Universitarias y de Investigación (AGAUR - 2023 BP 00141), Generalitat Valenciana (ASFAE/2022/008), Ministry of Science and Innovation (RYC2019-028510-I, RYC2020-030254-I, RYC2021-031273-I, RYC2022-038164-I), Ministerio de Ciencia, Innovación y Universidades/Agencia Estatal de Investigaci\'on (PID2022-142604OB-C22); Sweden: Carl Trygger Foundation (Carl Trygger Foundation CTS 22:2312), Swedish Research Council (Swedish Research Council 2023-04654, VR 2021-03651, VR 2022-03845, VR 2022-04683, VR 2023-03403, VR 2024-05451), Knut and Alice Wallenberg Foundation (KAW 2018.0458, KAW 2022.0358, KAW 2023.0366); Switzerland: Swiss National Science Foundation (SNSF - PCEFP2\_194658); United Kingdom: The Binks Trust, Royal Society (NIF-R1-231091); United States of America: U.S. Department of Energy (ECA DE-AC02-76SF00515), John Templeton Foundation (John Templeton Foundation 63206), Neubauer Family Foundation.

%
%


%
%
\printbibliography
\clearpage
 
\begin{flushleft}
\hypersetup{urlcolor=black}
{\Large The ATLAS Collaboration}

\bigskip

\AtlasOrcid[0000-0002-6665-4934]{G.~Aad}$^\textrm{\scriptsize 103}$,
\AtlasOrcid[0000-0001-7616-1554]{E.~Aakvaag}$^\textrm{\scriptsize 17}$,
\AtlasOrcid[0000-0002-5888-2734]{B.~Abbott}$^\textrm{\scriptsize 122}$,
\AtlasOrcid[0000-0002-0287-5869]{S.~Abdelhameed}$^\textrm{\scriptsize 118a}$,
\AtlasOrcid[0000-0002-1002-1652]{K.~Abeling}$^\textrm{\scriptsize 55}$,
\AtlasOrcid[0000-0001-5763-2760]{N.J.~Abicht}$^\textrm{\scriptsize 49}$,
\AtlasOrcid[0000-0002-8496-9294]{S.H.~Abidi}$^\textrm{\scriptsize 30}$,
\AtlasOrcid[0009-0003-6578-220X]{M.~Aboelela}$^\textrm{\scriptsize 45}$,
\AtlasOrcid[0000-0002-9987-2292]{A.~Aboulhorma}$^\textrm{\scriptsize 36e}$,
\AtlasOrcid[0000-0001-5329-6640]{H.~Abramowicz}$^\textrm{\scriptsize 156}$
% ...
\bigskip
\\

% ...
$^{17}$Department for Physics and Technology, University of Bergen, Bergen; Norway.\\
% ...
$^{30}$Physics Department, Brookhaven National Laboratory, Upton NY; United States of America.\\
% ...
$^{36}$$^{(a)}$Facult\'e des Sciences Ain Chock, Universit\'e Hassan II de Casablanca;$^{(b)}$Facult\'{e} des Sciences, Universit\'{e} Ibn-Tofail, K\'{e}nitra;$^{(c)}$Facult\'e des Sciences Semlalia, Universit\'e Cadi Ayyad, LPHEA-Marrakech;$^{(d)}$LPMR, Facult\'e des Sciences, Universit\'e Mohamed Premier, Oujda;$^{(e)}$Facult\'e des sciences, Universit\'e Mohammed V, Rabat;$^{(f)}$Institute of Applied Physics, Mohammed VI Polytechnic University, Ben Guerir; Morocco.\\
% ...
$^{45}$Physics Department, Southern Methodist University, Dallas TX; United States of America.\\
% ...
$^{49}$Fakult\"{a}t Physik, Technische Universit{\"a}t Dortmund, Dortmund; Germany.\\
% ...
$^{55}$II. Physikalisches Institut, Georg-August-Universit\"{a}t G\"ottingen, G\"ottingen; Germany.\\
% ...
$^{103}$CPPM, Aix-Marseille Universit\'e, CNRS/IN2P3, Marseille; France.\\
% ...
$^{118}$$^{(a)}$New York University Abu Dhabi, Abu Dhabi;$^{(b)}$United Arab Emirates University, Al Ain; United Arab Emirates.\\
% ...
$^{122}$Homer L. Dodge Department of Physics and Astronomy, University of Oklahoma, Norman OK; United States of America.\\
% ...
$^{156}$Raymond and Beverly Sackler School of Physics and Astronomy, Tel Aviv University, Tel Aviv; Israel.\\
% ...
$^{a}$ Also at Affiliated with an institute formerly covered by a cooperation agreement with CERN.\\
$^{b}$ Also at An-Najah National University, Nablus; Palestine.\\
$^{c}$ Also at Borough of Manhattan Community College, City University of New York, New York NY; United States of America.\\
$^{d}$ Also at Center for Interdisciplinary Research and Innovation (CIRI-AUTH), Thessaloniki; Greece.\\
$^{e}$ Also at Centre of Physics of the Universities of Minho and Porto (CF-UM-UP); Portugal.\\
$^{f}$ Also at CERN, Geneva; Switzerland.\\
% ...

\end{flushleft}


%

%
%

%

%
%
%
%

\end{document}